\documentclass[aps,prl,superscriptaddress,twocolumn]{revtex4}
\usepackage{amsmath}
\usepackage{amsfonts}
\usepackage{ulem}
\usepackage{color}
\usepackage{float}
\usepackage{soul}
\usepackage{graphicx}
\usepackage{amssymb}
\usepackage{epstopdf}

\usepackage{setspace}

\newcommand{\be}{\begin{equation}}
\newcommand{\ee}{\end{equation}}
\newcommand{\ber}{\begin{eqnarray}}
\newcommand{\eer}{\end{eqnarray}}
\newcommand{\nn}{\nonumber}

\begin{document}
\title{Hot-Carrier Cooling in High-Quality Graphene is Intrinsically Limited by Optical Phonons}

\author{Eva A. A. Pogna$^{1,2}$, Xiaoyu Jia$^{3}$, Alessandro Principi$^{4}$, Alexander Block$^{5}$, Luca Banszerus$^{6}$, Jincan Zhang$^{7,8}$, Xiaoting Liu$^{7,8}$, Thibault Sohier$^{9}$, Stiven Forti$^{10}$, Karuppasamy Soundarapandian$^{11}$, Bernat Terr\'es$^{11}$,Jake D. Mehew$^{5}$, Chiara Trovatello$^{2}$, Camilla Coletti$^{10,12}$, Frank H.L. Koppens$^{11,13}$, Mischa Bonn$^{3}$, Niek van Hulst$^{11,13}$, Matthieu J. Verstraete$^{9}$, Hailin Peng$^{7,8}$, Zhongfan Liu$^{7,8}$, Christoph Stampfer$^{6}$, Giulio Cerullo$^{2}$, Klaas-Jan Tielrooij$^{5,*}$
}

\affiliation{NEST, Istituto Nanoscienze-CNR and Scuola Normale Superiore, 56127 Pisa, Italy\\
$^{2}$Department of Physics, Politecnico di Milano, 20133 Milan, Italy\\
$^{3}$Max-Planck-Institut f{\"u}r Polymerforschung, 55128 Mainz, Germany\\
$^{4}$School of Physics and Astronomy, University of Manchester, Manchester, UK\\  
$^{5}$Catalan Institute of Nanoscience and Nanotechnology (ICN2), BIST \& CSIC, Campus UAB, Bellaterra (Barcelona), 08193, Spain\\
$^{6}$JARA-FIT and 2nd Institute of Physics, RWTH Aachen University, 52074 Aachen, Germany, EU\\
$^{7}$Center for Nanochemistry, College of Chemistry and Molecular Engineering, Academy for Advanced Interdisciplinary Studies, Peking University, Beijing 100871, China\\
$^{8}$Beijing Graphene Institute, Beijing 100095, P. R. China\\
$^{9}$NanoMat/Q-Mat/CESAM, Universit{\'e} de Li{\`e}ge (B5), B-4000 Li{\`e}ge, Belgium\\
$^{10}$Center for Nanotechnology Innovation IIT@NEST, Piazza San Silvestro 12, 56127 Pisa, Italy\\
$^{11}$ICFO - Institut de Ci\`{e}ncies Fot\`{o}niques, BIST, Castelldefels (Barcelona) 08860, Spain\\
$^{12}$Graphene Labs, Via Morego 30, 16163 Genova, Italy\\
$^{13}$ICREA - Instituci\'o Catalana de Re\c{c}erca i Estudis Avancats, 08010 Barcelona, Spain\\
$^{*}$klaas.tielrooij@icn2.cat} 

%\keywords{graphene, cooling dynamics, hot electrons, transient absorption microscopy, optical phonons, phonon bottleneck}

\begin{abstract}
Many promising optoelectronic devices, such as broadband photodetectors, nonlinear frequency converters, and building blocks for data communication systems, exploit photoexcited charge carriers in graphene. For these systems, it is essential to understand, and eventually control, the cooling dynamics of the photoinduced hot-carrier distribution. There is, however, still an active debate on the different mechanisms that contribute to hot-carrier cooling. In particular, the intrinsic cooling mechanism that ultimately limits the cooling dynamics remains an open question. Here, we address this question by studying two technologically relevant systems, consisting of high-quality graphene with a mobility $>$10,000 cm$^2$V$^{-1}$s$^{-1}$ and environments that do not efficiently take up electronic heat from graphene: WSe$_2$-encapsulated graphene and suspended graphene. We study the cooling dynamics of these two high-quality graphene systems using ultrafast pump-probe spectroscopy at room temperature. Cooling via disorder-assisted acoustic phonon scattering and out-of-plane heat transfer to the environment is relatively inefficient in these systems, predicting a cooling time of tens of picoseconds. However, we observe much faster cooling, on a timescale of a few picoseconds. We attribute this to an intrinsic cooling mechanism, where carriers in the hot-carrier distribution with enough kinetic energy emit optical phonons. During phonon emission, the electronic system continuously re-thermalizes, re-creating carriers with enough energy to emit optical phonons. We develop an analytical model that explains the observed dynamics, where cooling is eventually limited by optical-to-acoustic phonon coupling. These fundamental insights into the intrinsic cooling mechanism of hot carriers in graphene will play a key role in guiding the development of graphene-based optoelectronic devices.
\end{abstract}
\maketitle
\section{Introduction}
The ultrafast dynamics of photoexcited charge carriers in graphene have received ample attention, initially driven by fundamental scientific interest in the intriguing electron-electron and electron-phonon interactions, \textit{cf.}\ Refs.\cite{George2008,Breusing2011a,Brida2013,Gierz2013,Tielrooij2013}. More recently, interest has multiplied as a result of the emergence of highly promising technological applications that exploit these ultrafast dynamics. One example is ultrafast photodetection of visible (VIS) and infrared light \cite{Xia2009,Koppens2014}, and even terahertz (THz) radiation \cite{Bandurin2018,Castilla2019,Viti2020}. The ultrafast electronic response of graphene to incoming light has also led to the development of several concepts with relevance for data communication technologies, including modulators and receivers \cite{Liu2011,Romagnoli2018,Muench2019}, and was demonstrated to be crucial for tailoring nonlinear photonics applications \cite{Hafez2018,Soavi2018,Soavi2019,Deinert2020}. Part of the aforementioned applications, for instance photodetectors and receivers, exploit the ultrafast photo-thermoelectric effect in graphene \cite{Gabor2011,Tielrooij2015}, where a longer hot-carrier cooling time leads to an increased photoresponse. Other applications, such as modulators, could instead benefit from a short cooling time, as this can lead to a higher modulation speed, \textit{i.e.}\ to a larger bandwidth. Generally, these applications require high-quality graphene with a high electrical mobility. Clearly, it is crucial to properly understand the cooling dynamics of hot carriers in graphene, and in particular to identify the intrinsic mechanism that ultimately determines the cooling process in high-quality systems.
	
The decay of photoexcited charges in graphene occurs through a variety of dynamical processes that are the result of the specific properties of Dirac electrons and graphene phonons. Photo-excited charge carriers in graphene first undergo thermalization on a $<$100 femtosecond timescale, leading to a state with an elevated carrier temperature, \textit{i.e.}\ a broadened Fermi-Dirac distribution \cite{Breusing2011a,Brida2013,Gierz2013}. This heating process is very efficient \cite{Tielrooij2013,Tomadin2018}, and due to the small electronic heat capacity of graphene \cite{Fong2013}, the electron temperature $T_{\rm e}$ can be increased significantly ($T_{\rm e}>>$ 500 K). The hot-carrier state then relaxes back to the ground state with the electronic system at ambient temperature. This cooling process occurs through the interaction between charge carriers and graphene optical phonons, graphene acoustic phonons, as well as substrate phonons in nearby materials. These phonon-induced relaxation mechanisms have intricate dependencies on intrinsic (sample-dependent) parameters (such as disorder density and intrinsic doping), as well as extrinsic experimental parameters (such as photon energy, incident fluence, and ambient temperature). Thus, determining the dominant cooling channel(s) for excited graphene charge carriers has been challenging, and is still subject of debate. 
	
It is generally believed that charge carriers with enough kinetic energy can relax on an ultrafast timescale of a few hundred femtoseconds by interacting with strongly-coupled optical phonons \cite{Kampfrath2005,Breusing2011a,Hale2011}. These are the optical phonons at the $\Gamma$ and $K$ points with an energy of 0.2 eV and 0.16 eV, respectively \cite{Mounet2005}. Thus, charge carriers with an energy $>$0.16 eV above the chemical potential can efficiently cool by optical phonon emission \cite{Mihnev2016a}. Carriers with an energy $<$0.16 eV can only couple to acoustic phonons, generally resulting in very inefficient cooling with decay times up to the nanosecond range \cite{macdonald2009}. In the case of graphene with significant disorder, however, coupling to acoustic phonons becomes much more efficient through disorder-assisted scattering. By scattering with defects, the large momentum mismatch between electrons and acoustic phonons is overcome \cite{Song2012a}. This disorder-assisted ``supercollision'' cooling process leads to typical cooling times of a few picoseconds at room temperature, as measured by electrical \cite{Betz2012b}, optical \cite{Graham2013a,Alencar2014} and optoelectronic \cite{Graham2013} techniques. Since disorder also limits the electrical mobility, this disorder-assisted cooling mechanism tends to play an increasingly important role for graphene with lower electrical mobility, typically below $\sim$10,000 cm$^2$V$^{-1}$s$^{-1}$.  
	
In the case of graphene with a mobility above 10,000 cm$^2$V$^{-1}$s$^{-1}$, which we here refer to as ``high-quality graphene'', the disorder density is low enough that disorder-assisted cooling will likely not play an important role. Recent experiments on high-quality hBN-encapsulated graphene with a mobility above 30,000 cm$^2$V$^{-1}$s$^{-1}$, however, did not show a strong increase in cooling time \cite{Tielrooij2018}. This is because a parallel relaxation mechanism emerges in these heterostructures: hot carriers in graphene can  decay through near-field interaction with hyperbolic phonon polaritons of the hBN \cite{Principi2017,Yang2018,Tielrooij2018}. Hyperbolic phonon polaritons occur in spectral regions, where the in- and out-of-plane permittivities ($\epsilon_{||}$ and $\epsilon_{\perp}$, respectively) have opposite signs, \textit{i.e.}\ the permittivity product $\epsilon_{||} \cdot \epsilon_{\perp}$ is negative \cite{Caldwell2014}. These hyperbolic spectral regions contain a high density of optical modes, which are accessible via near-field interaction with a large range of momentum vectors. As a result,  out-of-plane super-Planckian cooling for hBN-encapsulated graphene occurs with a decay time of a few picoseconds at room temperature \cite{Principi2017,Yang2018,Tielrooij2018}. The important open question therefore remains: what is the \textit{intrinsic} physical mechanism that ultimately limits the cooling of hot carriers in high-quality graphene?

Here, we address this question using time-resolved optical measurements of the cooling dynamics in two specifically chosen material systems: WSe$_2$-encapsulated graphene and suspended graphene. Both systems contain ``high-quality'' graphene, according to our definition of having a mobility $>$10,000 cm$^2$V$^{-1}$s$^{-1}$, which is high enough to make ``supercollision'' cooling inefficient. Furthermore, they have basically non-hyperbolic environments, such that out-of-plane cooling to hyperbolic phonons does not play any role. We show that, despite eliminating these two relaxation channels, cooling still happens on a timescale of a few picoseconds. We attribute this to the high-energy tail of the hot-carrier distribution, where electrons with an energy $>$0.16 eV above the chemical potential reside. These electrons lose energy by coupling to optical phonons, which in turn couple to graphene acoustic phonons, while the electronic system continuously re-thermalizes. Probing the transient optical properties of graphene in the VIS, near-infrared (NIR) and THz ranges, we will show that this cooling mechanism is consistent with the experimentally obtained cooling dynamics. 
\section{Results}
\subsection{Cooling dynamics in high-quality WSe$_2$-encapsulated graphene}
\begin{figure*}[ht!p]
	\centerline{\includegraphics[scale=0.92]{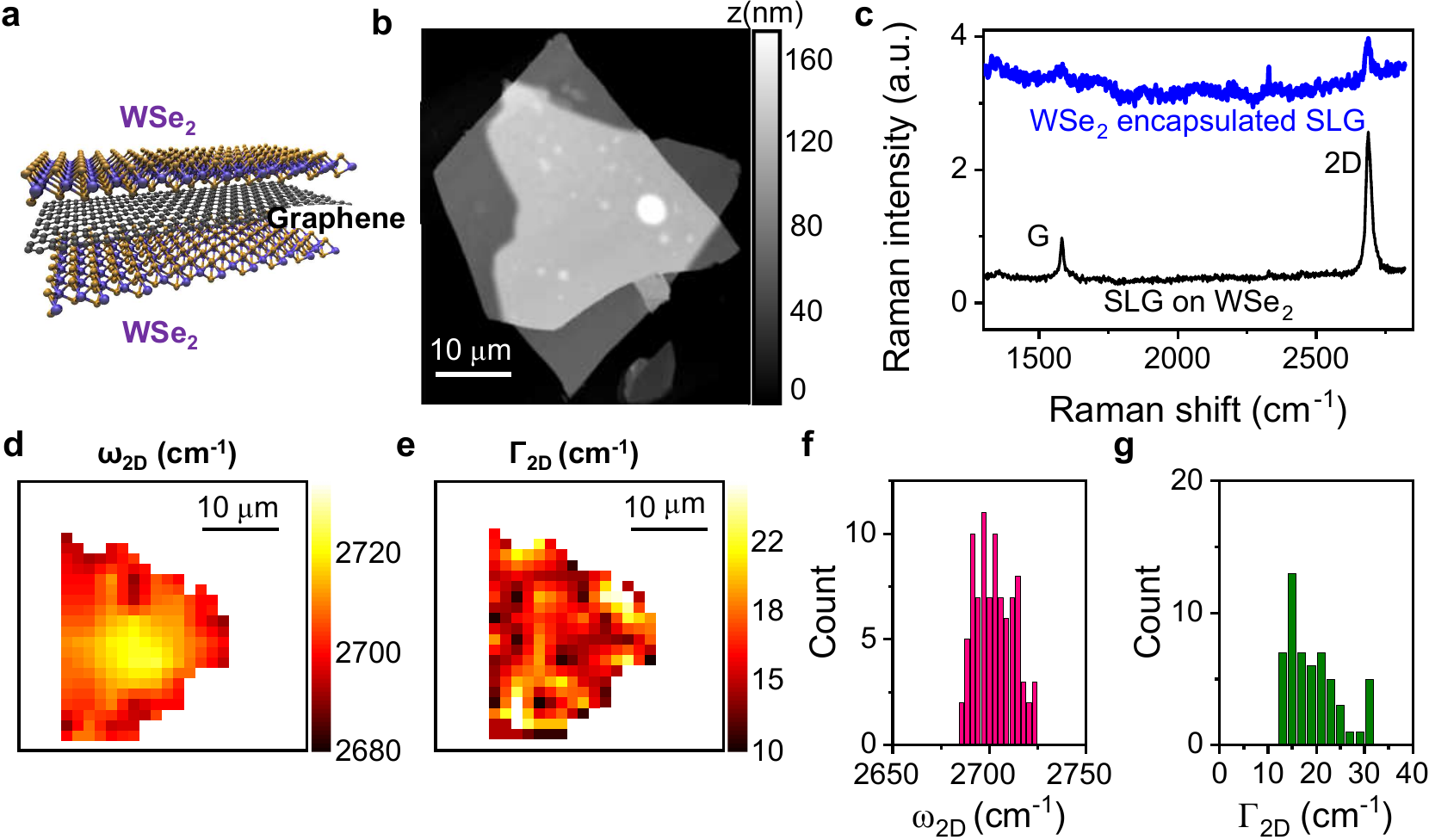}}
	\caption{\textbf{Characterization of the WSe$_2$-encapsulated graphene sample.} \textbf{a)} Sketch of the encapsulated graphene prepared by exfoliation and dry transfer of two  flakes of WSe$_2$ and monolayer graphene \cite{Backes2020}. \textbf{b)} Atomic force microscopy image of the sample, used to determine the thicknesses of the WSe$_2$ flakes. \textbf{c)} Raman spectra of fully encapsulated graphene (blue line) and of semi-encapsulated graphene (black line) obtained with 532 nm laser source. Encapsulation gives rise to a background, reduced graphene Raman signatures' (G and 2D peaks) intensity, and a clear blue-shift of the 2D peak. \textbf{d-e)} Maps of the 2D peak frequency $\omega_{\rm 2D}$ \textbf{(d)} and width $\Gamma_{\rm 2D}$ \textbf{(e)} extracted from Raman spectra at different positions on the sample fitted with a Voigt function centered in $\omega_{\rm 2D}$. 
		\textbf{f-g)} Statistical distribution of $\omega_{\rm 2D}$ \textbf{(f)} and $\Gamma_{\rm 2D}$ \textbf{(g)} in the encapsulated graphene obtained from the maps in panels d-e. The narrow width suggests a high electrical mobility. 
	}\label{f:raman}
\end{figure*}

The first material system we study is WSe$_2$-encapsulated graphene (see Fig.\ \ref{f:raman}a). Encapsulation with materials offering atomically planar surfaces via van der Waals stacking, is a successful route for obtaining high-quality graphene, suppressing rippling, preserving high carrier mobility\cite{Dean2010} and partially screening long-distance Coulomb scattering with substrate charges\cite{wang2013one}. 

Besides hBN, WSe$_2$ is arguably one of the most promising encapsulation materials for various technological applications of graphene\cite{Banszerus2017}. One of the main reasons is that it leads to a very high room-temperature mobility, which in a recent study reached up to 350,000 cm$^2$V$^{-1}$s$^{-1}$ for hBN/graphene/WSe$_2$ \cite{Banszerus2019}. Our sample of WSe$_2$-encapsulated monolayer graphene was prepared using exfoliation and dry transfer. The WSe$_2$ flakes were obtained by adhesive tape exfoliation of WSe$_2$ crystals in the trigonal prismatic phase synthesized by chemical vapor transport, as detailed in the Methods. Bottom and top WSe$_2$ flakes are 63 and 61 nm thick, respectively, as measured by atomic force microscopy (see Fig.\ \ref{f:raman}b and Supporting Informations, SI). The WSe$_2$/graphene/WSe$_2$ heterostructure was transferred onto a 280 $\mu$m thick CaF$_2$ substrate which is transparent for VIS and NIR light and allows for pump-probe measurements in transmission geometry. 
\newline We first characterize the WSe$_2$-encapsulated graphene using Raman spectroscopy with a 532 nm laser source (see Methods for details), in order to assess the quality of graphene and extract an estimation of its Fermi energy and charge mobility. In the regions where graphene is semi-encapsulated, the Raman fingerprints of graphene, the G-peak and 2D-peak, are clearly visible (see Fig.\ \ref{f:raman}c). In the fully encapsulated region, the spectrum is dominated by a large background. Nevertheless, by extending the exposure time, we can identify the G-peak with center frequency of $\omega_{\rm G}\sim$1584 cm$^{-1}$ and width of $\Gamma_{\rm G}\sim$20 cm$^{-1}$ (spectra in the SI). From the G-peak position and width we estimate that $|E_{\rm F}|\leq$ 0.1 eV, in agreement with a previous observation for graphene supported by single-layer WSe$_2$\cite{Banszerus2019}. Examining the 2D-peak at many positions in the encapsulated portion of graphene, we find that it is blue-shifted from $\sim$2689 cm$^{-1}$ to $\sim$2710 cm$^{-1}$ after encapsulation, see Fig.\ \ref{f:raman}d. The distribution of the 2D-peak widths $\Gamma_{\rm 2D}$ in Fig.\ \ref{f:raman} is centered at 18 cm$^{-1}$, indicating that WSe$_2$-encapsulated graphene has a rather homogeneous strain distribution, with little intra-valley scattering and high mobility\cite{Neumann2016}. Based on the empirical correlation of Ref.~\citenum{Robinson2009}, the measured $\Gamma_{\rm 2D}$ corresponds to a mobility of $\sim$80,000 cm$^2$V$^{-1}$s$^{-1}$. In order to verify this estimate, we measured the transport properties at room temperature of a similar WSe$_2$-encapsulated graphene heterostructure with two metal contacts and a backgate. We obtained an electron (hole) mobility of $\sim$39,000 cm$^{2}$V$^{-1}$s$^{-1}$ ($\sim$36,000 cm$^{2}$V$^{-1}$s$^{-1}$, see SI). 

\begin{figure*} [ht!p]
	\centerline{\includegraphics[width=\textwidth]{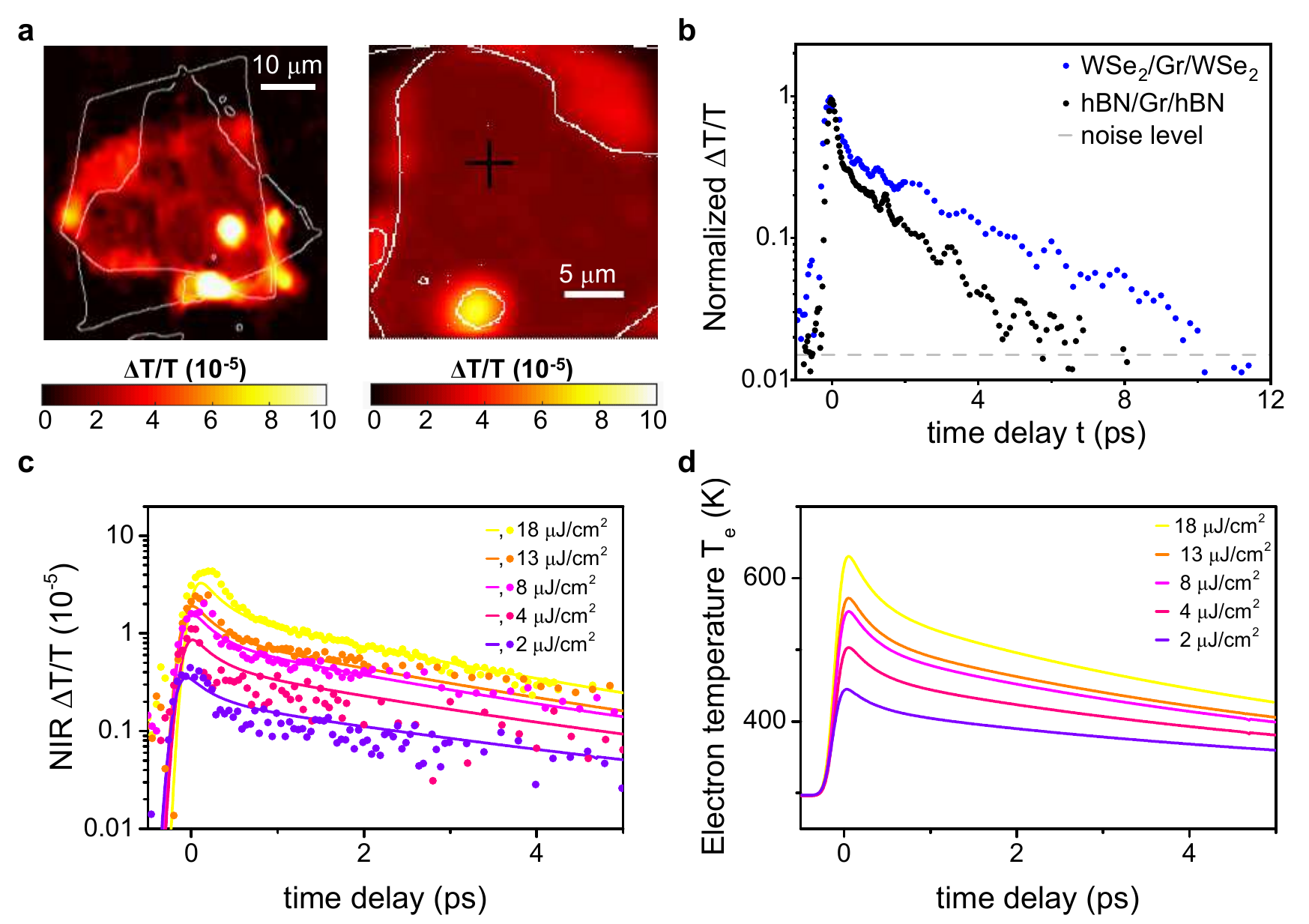}}
	\caption{ \textbf{Hot-carrier cooling dynamics of WSe$_2$-encapsulated graphene.} \textbf{a)} Transient transmission $\Delta T/T$ maps at a fixed time delay $t$ = 100 fs between pump and probe pulses acquired with pump at 0.8 eV (1550 nm) and probe at 0.729 eV (1700 nm). White lines indicate the edges of WSe$_2$ flakes extracted from an optical image. The panel on the right shows a zoom of the left panel, and the black cross indicates the location in the fully-encapsulated region, where the $\Delta T/T$ dynamics are measured. \textbf{b)} Comparison of normalized $\Delta T/T$ relaxation dynamics of graphene encapsulated by WSe$_2$ (blue dots) and hBN (black dots), using a pump fluence of $\sim$10 $\mu$J/cm$^2$ in both cases. Encapsulation with hBN gives rise to faster dynamics, which we attribute to out-of-plane cooling to hyperbolic phonons of hBN \cite{Principi2017,Tielrooij2018,Yang2018}. This cooling mechanism is much less efficient for WSe$_2$ encapsulation.  \textbf{c)} $\Delta T/T$ dynamics (dots) for WSe$_2$-encapsulated graphene at five different pump fluences from 2 to 18 $\mu$J/cm$^{2}$. The positive  $\Delta T/T$ is due to pump-induced heating, leading to Pauli blocking of probe interband transitions. The solid lines are the calculated $\Delta T/T$ dynamics, based on the intrinsic hot-electron cooling mechanism via the combination of optical-phonon emission, continuous re-thermalization of the electron distribution, and coupling of optical to acoustic phonons. \textbf{d)} Dynamical evolution of the electron temperature $T_{\rm e}$, corresponding to the calculated $\Delta T/T$ dynamics in panel c. Here, fast decay due to the electron-optical-phonon coupling is followed by slower decay via optical-to-acoustic phonon coupling. The former component shows up very strongly in the TA signal as a consequence of the strongly superlinear relation between the TA signal $\Delta T/T$ and the change in carrier temperature $\Delta T_{\rm e}$.}
	\label{f:Encdyn}
\end{figure*}
	
With such high mobility values, disorder-assisted ``supercollision'' cooling predicts a characteristic cooling timescale of $>$10 ps (see SI). 	
Having established the high quality of our WSe$_2$-encapsulated graphene sample, we now study its hot-carrier cooling dynamics. We perform time-resolved transient absorption (TA) measurements in the NIR range using a high-sensitivity microscope (see Methods for details on the experimental setup), which effectively serves as an ultrafast electronic thermometer. All experiments are performed under ambient conditions and at room temperature, monitoring the time-dependent differential transmission $\Delta T/T$ $(t)$. We use pump pulses at 1550 nm (0.8 eV) with a 40 MHz repetition rate, and probe pulses at 1700 nm (0.729 eV). These photon energies are chosen well below the band gap of WSe$_2$ ($\sim$1.35 eV for multilayer WSe$_2$), to avoid photoexcitation of charges in the encapsulant (see reflectance measurements in the SI). The pump pulses are absorbed in graphene, where they induce carrier heating, \textit{i.e.}\ a broadening of the Fermi-Dirac distribution of the electronic system. The probe pulses are sensitive to this, because the broadening leads to Pauli blocking of the interband transitions at the probe photon energy. Therefore, the pump-induced increase in probe transmission is a measure of the electronic temperature. Accordingly, the $\Delta T/T$ signal in Fig.\ \ref{f:Encdyn}a is positive, indicating pump-induced transmission increase, \textit{i.e.}\ absorption bleaching. We can distinguish between fully encapsulated and semi-encapsulated graphene regions in Fig.\ \ref{f:Encdyn}a, as the former gives rise to a lower transient signal (at a time delay of $\sim$100 fs), which is likely the result of strong Fresnel reflections at the air-WSe$_2$ and WSe$_2$-graphene interfaces. 

Figure \ref{f:Encdyn}b shows the relaxation dynamics of the $\Delta T/T$ signal of WSe$_2$-encapsulated graphene, and compares it to the dynamics for hBN-encapsulated graphene. The decay in WSe$_2$-encapsulated graphene is significantly slower: the signal decays to 10\% of the initial value (at time zero) after a time delay of $\sim$5.5 ps for WSe$_2$-encapsulated graphene, instead of a delay of $\sim$2 ps for the hBN-encapsulated graphene. As shown recently, in high-quality, hBN-encapsulated graphene, hot-carrier cooling is dominated by out-of-plane coupling to hyperbolic phonons in hBN \cite{Principi2017,Tielrooij2018,Yang2018}. Since the hyperbolic nature of hBN is crucial for making this cooling channel efficient \cite{Principi2017}, we study the hyperbolicity of WSe$_2$ using \textit{ab-initio} density functional theory calculations. The comparison of in-plane and out-of-plane permittivities of WSe$_2$ and hBN, reported in the SI, shows that WSe$_2$ is much less hyperbolic than hBN. Therefore, out-of-plane cooling to the  phonons in the encapsulant will be relatively inefficient in WSe$_2$-encapsulated graphene, in qualitative agreement with the slower cooling that we observe. Quantitatively, however, both disorder-assisted ``supercollision'' cooling and out-of-plane cooling to the encapsulant would predict longer cooling times (see SI), well above 10 ps for the WSe$_2$-encapsulated graphene sample, which is not what we observe.

In order to explain the relatively fast cooling dynamics observed in the WSe$_2$-encapsulated graphene, we therefore consider an intrinsic cooling mechanism, based on the coupling of electrons with energy $>$0.16 eV above the chemical potential to optical phonons. The optical phonons, at the $\Gamma$ and $K$ points, in turn, are anharmonically coupled to acoustic phonons. An important ingredient for this cooling channel is the continuous re-thermalization of the electronic system. Microscopically, this means that once the electrons with  high enough energy (more than 0.16 eV above the chemical potential) have relaxed by coupling to optical phonons, the remaining electrons will re-thermalize. This means that some electrons will end up with an energy that is high enough to emit optical phonons. Even at an electron temperature of 300 K, there is a significant fraction of electrons that can emit optical phonons. As a result, this is a rather efficient cooling channel for graphene at room temperature. We will describe the analytical model of this cooling mechanism in more detail in the Discussion and SI. 

We calculate the cooling dynamics and compare the results to the experimental data. In the calculation, we also take into account the superlinear relation connecting the $\Delta T/T$ signal to the change in carrier temperature. As shown in Fig.\ \ref{f:Encdyn}c, the experimental data and calculated signal are in good agreement when we use an optical-to-acoustic phonon decay time of 2 ps and a Fermi energy of 0.1 eV. This optical phonon lifetime lies within the range of 1.2 -- 2.55 ps, previously reported \cite{Kang2010,Lui2010,Wang2010,Wu2012,Bonini2007} for graphene. Figure \ref{f:Encdyn}d shows the temperature dynamics corresponding to the calculated transient signals in Fig.\ \ref{f:Encdyn}c, which are clearly non-exponential. The fast initial decay corresponds to efficient coupling to optical phonons, whereas the slower subsequent decay is the result of the hot-phonon bottleneck, where optical phonons are cooling to acoustic phonons. 
\\
	
\begin{figure*} [htp!]
	\centerline{\includegraphics[width=\textwidth]{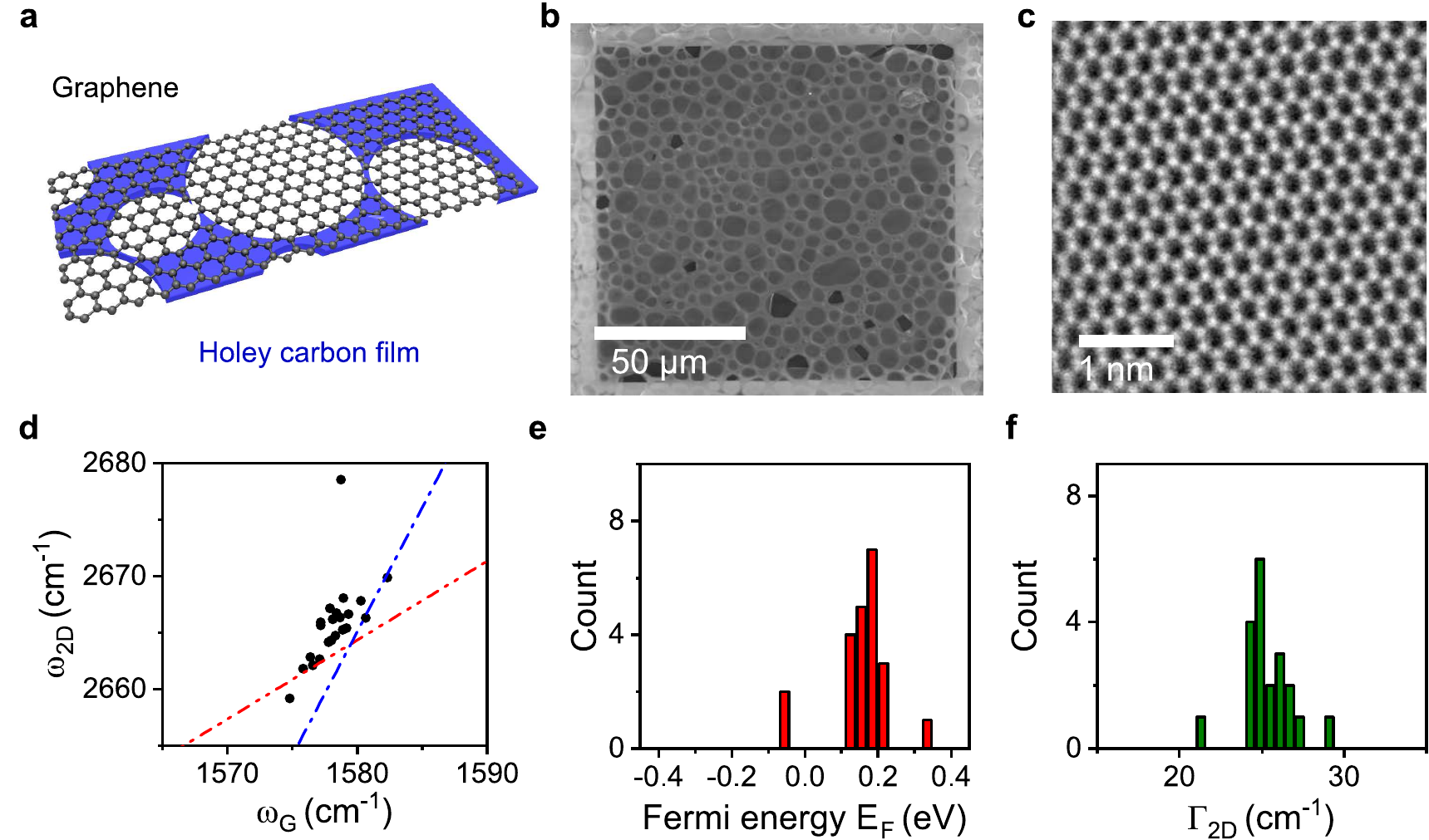}}
	\caption{ \textbf{Characterization of suspended high-quality graphene.}
		\textbf{a)} Sketch of graphene suspended on a holey carbon film. \textbf{b)} Scanning electron microscopy images of suspended graphene on holey carbon, showing relatively large hole sizes and high yield. \textbf{c)} TEM image of the suspended graphene, showing absence of disorder. \textbf{d)} Plot of 2D peak frequency $\omega_{\rm 2D}$ as a function of G peak frequency $\omega_{G}$. Dashed lines represent the expected  dependence in the strain-free (red dot-dot-dashed line) and doping-free (blue dot-dashed line) material. \textbf{e)} Fermi energy $E_{\rm F}$ distribution extracted from vector decomposition of peaks positions in panel d \cite{lee2012}, and centred at $|E_{\rm F}|\approx$0.18 eV. \textbf{f)} Obtained distribution of 2D-peak widths, indicating a mobility of $>$17,000 cm$^2$V$^{-1}$s$^{-1}$, following calculations of Ref.~\citenum{Robinson2009}.}
	\label{f:raman2}
\end{figure*}
\subsection{Cooling dynamics in high-quality suspended graphene}
In order to determine if this cooling mechanism is intrinsic to graphene, we study a second technologically relevant material system in which any thermal exchange with the environment is inherently excluded: high-quality suspended graphene. This sample contains large-area graphene grown by chemical vapour deposition (CVD), which was then transferred onto a transmission electron microscopy (TEM) grid, see Fig.\ \ref{f:raman2}a. This was achieved through a polymer-free approach, using the method described in Ref.~\citenum{Zhang2017}. As a TEM grid we used holey carbon, which is convenient as it has locations with relatively large holes with a diameter exceeding 10 $\mu$m. It has been shown that the graphene in such samples has a mobility well over 10,000 cm$^2$V$^{-1}$s$^{-1}$ \cite{Lin2019}. Given the large area and high quality of graphene prepared by this fabrication approach, which is furthermore scalable, this is a highly promising material system for a broad range of electronic and optoelectronic applications.
\\

We characterize our suspended graphene sample using various microscopic techniques, see Fig. \ref{f:raman2}. First, we show an image taken using scanning electron microscopy, evidencing excellent graphene coverage of the holes and providing an indication of hole sizes (see Fig.\ \ref{f:raman2}b). TEM measurements furthermore show an atomically perfect lattice (see Fig.\ \ref{f:raman2}c). We then perform Raman spectroscopy, in order to estimate the Fermi energy and charge mobility of the suspended graphene. Performing a strain-doping analysis of the G-peak and 2D-peak, obtained from several positions on the sample (see Fig. \ref{f:raman2}d), we obtain a distribution of carrier densities (see Fig.\ \ref{f:raman2}e) which corresponds to an average Fermi energy of $\sim$0.18 eV. Using the measured width of the 2D-peaks in Fig.\ \ref{f:raman2}e and the empirical correlation of Ref.~\citenum{Robinson2009} , we extract a mobility of $>$17,000 cm$^2$V$^{-1}$s$^{-1}$. This result is in agreement with electrical measurements performed on a sample of suspended graphene, prepared in an identical way, see Ref.~\citenum{Lin2019}. The extracted mobility confirms the high quality of the sample, for which disorder-assisted "supercollision" cooling predicts a cooling time $>$10 ps (see SI).
\\
	
\begin{figure*}
	\centerline{\includegraphics[width=\textwidth]{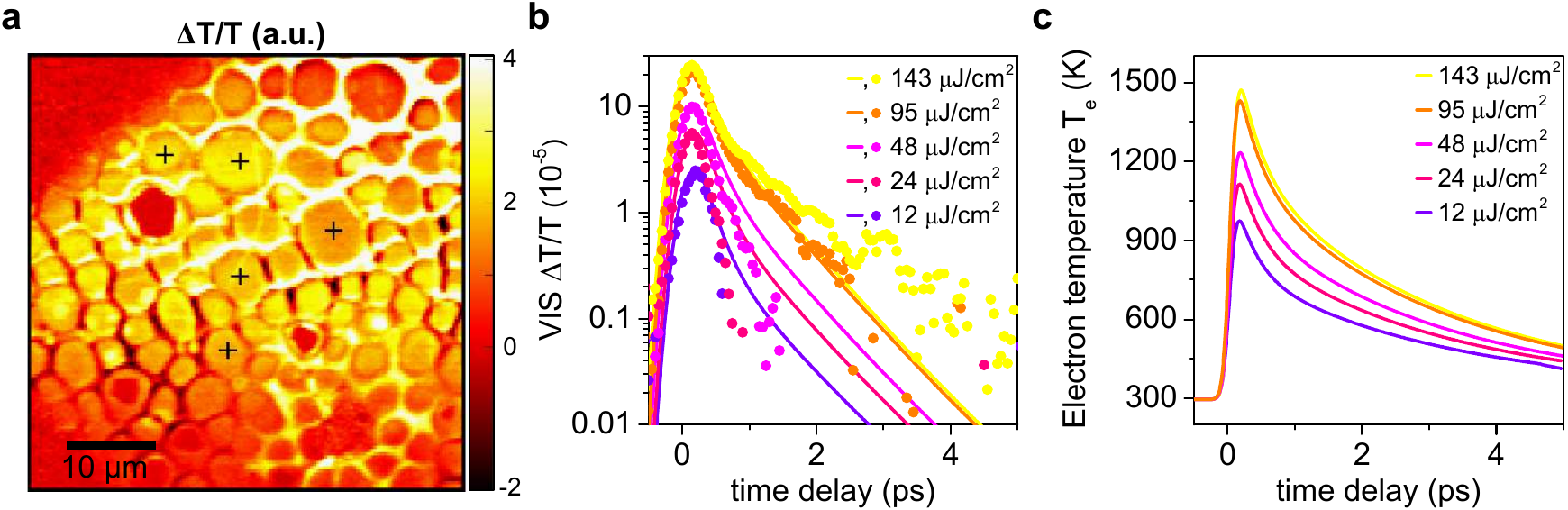}}
	\caption{\textbf{Hot-carrier cooling dynamics of suspended graphene.} \textbf{a)} Transient transmission $\Delta T/T$ map at time zero, showing individual holes where graphene is suspended, acquired with pump tuned at 3.1 eV (400 nm) and probe at 1.55 eV (800 nm).\textbf{b)} $\Delta T/T$ dynamics (coloured dots) for suspended graphene at five different pump fluences from 12 to 143 $\mu$J/cm$^{2}$. The positive $\Delta T/T$ results from  the pump-induced carrier heating that leads to decreased interband conductivity due to Pauli blocking, and thus decreased absorption of the NIR probe light. The coloured lines are the calculated $\Delta T/T$ dynamics, based on the intrinsic cooling mechanism where hot electrons cool via the combination of emission of optical phonons, continuous re-thermalization of the electron distribution, and coupling of optical to acoustic phonons. \textbf{c)} Dynamical evolution of the electron temperature $T_{\rm e}$, corresponding to the calculated $\Delta T/T$ dynamics in panel b. Here, fast decay due to electron-optical-phonon coupling is followed by slower decay due to optical-to-acoustic phonon coupling.}
	\label{Fig5}
\end{figure*}

Having established the high quality of our suspended graphene sample, we now study its hot-carrier cooling dynamics using time-resolved transient absorption microscopy (see Methods for details on the experimental setup). We use a pump tuned at 400 nm (3.1 eV) and a probe at 800 nm (1.55 eV). Similar to the NIR TA measurements, we observe photoinduced bleaching ($\Delta T/T>0$) of the probe, due to heating-induced Pauli blocking. The $\Delta T/T$ map at zero time delay (overlapping pump and probe pulses) in Fig.\
\ref{Fig5}a shows that we can resolve the individual holes spatially, and that a clearly distinctive transient signal comes from inside -- from the suspended graphene.
\\

The experimental TA dynamics of our high-quality suspended graphene sample are shown in Fig.\ \ref{Fig5}b for five different fluences. Similar transient optical responses were observed in suspended graphene prepared by exfoliation, in the VIS (400 nm pump, 800 nm probe) \cite{Malard2013}, and in the NIR (830 nm pump, 1100-1400 nm probe) \cite{Hale2011}. 
We describe the dynamics with the same cooling mechanism as for WSe$_2$-encapsulated graphene (see Discussion), with the same optical-to-acoustic phonon coupling time of 2 ps (details in the SI). For this sample we use an $E_{\rm F}$ = 0.15 eV,  close to the experimentally determined  average value. Figure \ref{Fig5}c reports the temperature dynamics corresponding to the $\Delta T/T$ dynamics in Fig.\ \ref{Fig5}b, again showing an initial fast decay due to coupling between electrons and optical phonons, followed by slower decay due to optical phonons coupling to acoustic phonons.

As a final experimental test, we complement our TA measurements -- which probe interband transitions -- with optical-pump THz probe (OPTP) measurements -- which probe intraband transitions, see Fig.\ S3 of SI. Despite the different optical transitions that are probed, both techniques essentially function as ultrafast electrical thermometers for graphene. This is because in both cases the probe absorption is affected by pump-induced changes in the carrier temperature. The observed OPTP dynamics, reported in the SI, can be described with the same model as before, confirming the validity of this intrinsic cooling mechanism.
\\
	
\section{Discussion}
	
\begin{figure*}[!ht]
	\centerline{\includegraphics[scale=1]{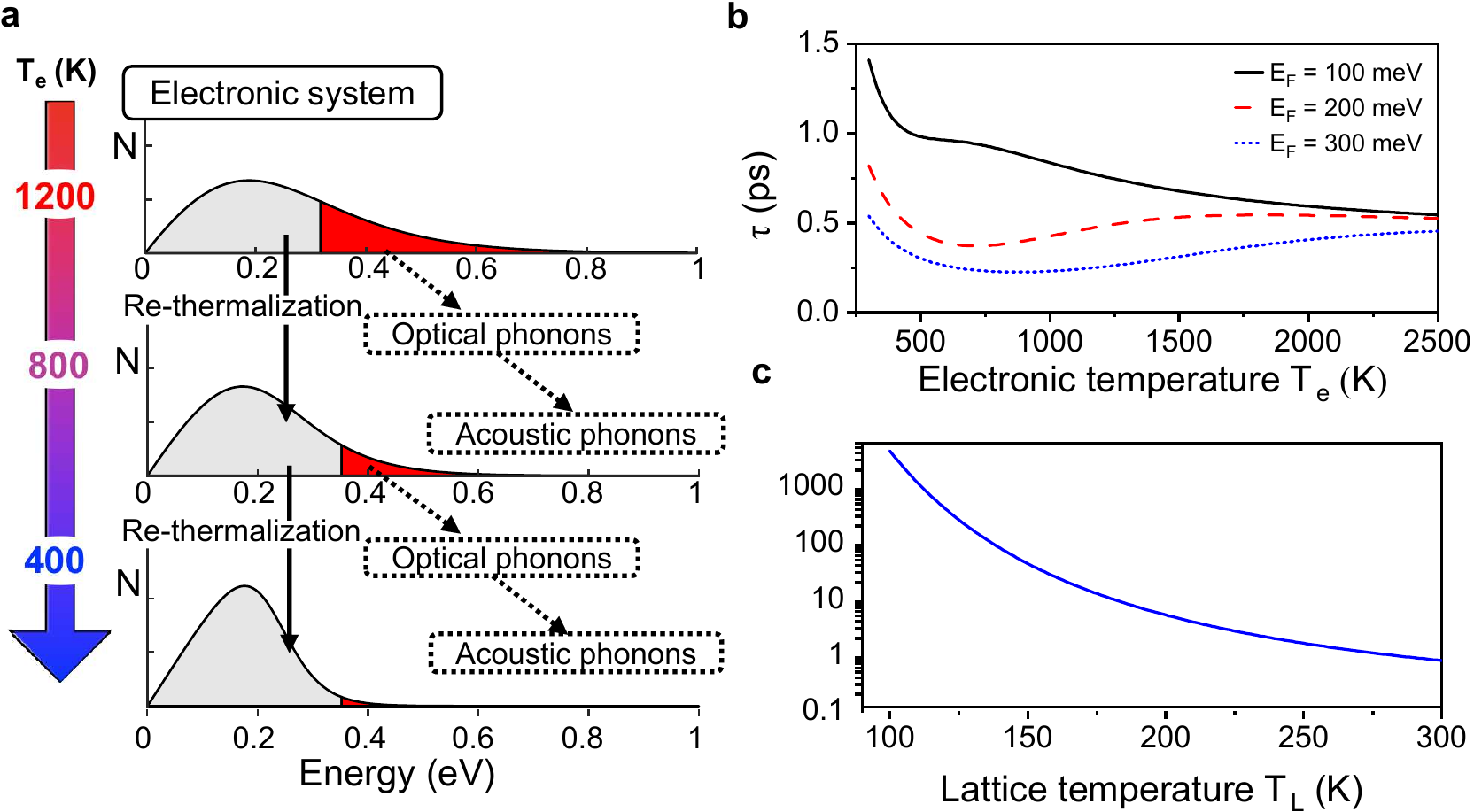}}
	\caption{\textbf{Hot-carrier cooling dynamics in high quality graphene.} \textbf{a)}. The process of electronic cooling explained through schematics of the carrier density as a function of carrier energy for three electron temperatures (1200, 800 and 400 K). Cooling occurs through a combination of \textit{i)} optical phonon emission by electrons with energy $>$0.16 eV above the chemical potential (red-shaded area) \textit{ii)} re-thermalization of the electronic distribution, and \textit{iii)} anharmonic coupling of optical phonons to acoustic phonons. \textbf{b)} Calculated "instantaneous" cooling time for a given initial electron temperature for three different Fermi energies.\textbf{c)} Calculated cooling time as a function of lattice temperature T$_L$ for a very small $\Delta T_{\rm e}$ and E$_{\rm F}$= 0.03 eV. For panels b-c, cooling occurs through optical phonon emission and continuous re-thermalization of the electronic system. The hot-phonon bottleneck is not included.} 
	\label{Fig1}
\end{figure*}

We now discuss in more detail the cooling mechanism that we have used to describe the experimentally obtained cooling dynamics (schematically shown in Fig.\ \ref{Fig1}a). First, we note that in many early time-resolved studies on graphene the dynamics were explained using a qualitatively similar mechanism involving electrons decaying to optical phonons, and a hot-phonon bottleneck, \textit{cf.}\ Refs.\ \cite{Wang2010,Hale2011,huang2011,Malard2013}. This mechanism, however, was thought to only mediate cooling for carriers with high enough energy to couple directly to opical phonons \cite{Viljas2010}. For the rest of the carriers in the hot-carrier distribution, alternative cooling channels were considered, such as disorder-assisted ``supercollision cooling'' to graphene acoustic phonons \cite{Song2012a,Graham2013,Betz2012b,Graham2013a} and out-of-plane cooling to (hyperbolic) substrate phonons \cite{Principi2017,Tielrooij2018,Yang2018}, which can both give rise to picosecond cooling at room temperature. In 2016, a microscopic, numerical, simulation of the cooling dynamics of hot carriers in graphene was presented, based on electron-to-optical phonon coupling, and including re-thermalization of the carrier system \cite{Mihnev2016a}. The calculated cooling times were used to explain qualitative trends in decay times measured by OPTP in samples of multilayer epitaxial graphene on SiC and monolayer graphene grown by CVD. These results motivated us to (re)consider cooling via optical phonons as the intrinsic cooling pathway for high-quality graphene, where disorder-assisted cooling and out-of-plane cooling are inefficient. The cooling mechanism is schematically explained in Fig. \ref{Fig1}a.
\\
	
We developed an analytical model to describe the hot-carrier cooling dynamics in graphene. The details of the derivation are shown in the SI. Briefly, we solve the following rate equations for the electron  temperature $T_e(t)$, and phonon temperature $T_\alpha(t)$:
\begin{equation}
\left\{
\begin{array}{l}
{\displaystyle C_e\big(T_e(t)\big) \partial_t T_e(t) = -\hbar \sum_{\alpha} \omega_\alpha {\cal R}_\alpha\big(T_e(t),T_\alpha(t)\big) }
\vspace{0.2cm}\\
{\displaystyle {\cal D}\big(\omega_\alpha,T_\alpha(t)\big) \partial_t T_\alpha(t) = \frac{{\cal R}_\alpha\big(T_e(t),T_\alpha(t)\big)}{M_\alpha\big(T_e(t)\big)}}\\
{\quad\quad\quad\quad\quad\quad\quad\quad -\gamma_\alpha\big[n_\alpha\big(T_\alpha(t)\big)-n_\alpha(T_\alpha^{(0)})\big] }
\end{array}
\right.
~
\end{equation}

The left-hand side of the first rate equation contains the electronic heat capacity $C_e\big(T_e(t)\big)$ and the temporal derivative of the electron temperature. The right-hand side describes the emission of optical phonons, where the sum is over the two optical phonon modes (labelled by $\alpha$), at the $\Gamma$ and $K$ point, $\omega_\alpha$ is the frequency of mode $\alpha$, and ${\cal R}_\alpha(T_e,T_\alpha)$ is the rate of $\alpha$-phonon emission. 

We calculate this rate analytically using a Boltzmann-equation approach, which contains one important input parameter, namely the electron-phonon coupling strength. For this, we use the value 11.4~eV/\AA{}, obtained by density functional theory calculations, and verified by experiments \cite{Sohier2014}. The left-hand side of the second rate equation contains ${\cal D}(\omega_\alpha,T_\alpha) = \partial n_\alpha(T_\alpha)/(\partial T_\alpha)$, where $n_\alpha(T_\alpha)$ is the phonon occupation function. The right-hand side contains a first term due to the emission of optical phonons by electrons, where the parameter $M_\alpha(T_e)$ measures the size of the portion (an annulus) of the phonon Brillouin zone that is heated in the electron-cooling process. This parameter depends, assuming the phonon dispersion to be flat, on the maximum momentum that can be exchanged between electrons and phonons, and thus on the electron temperature $T_e$. The second term describes the decay of optical phonons to acoustic phonons, where $T_\alpha^{(0)}$ is the equilibrium phonon temperature, {\it i.e.} the lattice temperature $T_{\rm L}$.  For the optical phonon decay term, we use the parameter $\gamma_\alpha$ as a phenomenological damping rate. There are essentially two adjustable  parameters in  our calculations: the optical  phonon lifetime $\gamma_\alpha^{-1}$, and the parameter $v$ that governs the temperature dependence of the phonon number density (see  SI). 

We have seen that this analytical model is able to accurately describe the experimentally obtained cooling dynamics. The hot-phonon bottleneck, occurring when the density of emitted optical phonons is so high that they cannot completely decay into acoustic phonons and part of their energy is scattered back to the electronic system, becomes more and more important with increasing initial $T_e$, and it leads to an overall slower cooling of the graphene hot carriers. For applications operating with a small heating $\Delta T_{\rm e}$, however, cooling is ultimately determined by electron-optical-phonon cooling. This regime is likely relevant for applications that require low power consumption with low light intensities. Therefore, we analytically study the cooling in the absence of the hot-phonon bottleneck, \textit{i.e.}\ when $\gamma_\alpha \rightarrow \infty$. We calculate that in this case, cooling at room temperature takes between $\sim$500 fs, for $E_{\rm F}$ = 0.3 eV, and $\sim$1.4 ps for 0.1 eV. Cooling to optical phonons will quickly become less efficient upon decreasing the lattice temperature $T_{\rm L}$. Around 200 K, we find a cooling time around 5 ps, whereas this increases to $\sim$ 4 ns at 100 K (see Fig.\ \ref{Fig1}c). We note that when $\Delta T_{\rm e}$ is not small, the effect of increased cooling time with decreased lattice temperature is much weaker. Thus, obtaining a longer intrinsic cooling time requires a reduction of both $T_{\rm L}$ and $\Delta T_{\rm e}$.
\\
	
\section{Conclusion}
	
Using three different time-resolved measurement techniques (NIR TA, VIS TA and OPTP) and two different high-quality, technologically relevant, graphene systems (WSe$_2$-encapsulated and suspended graphene), we have shown that hot carriers decay through an intrinsic mechanism involving optical phonon emission and constant re-thermalization of the electronic system. Electrons with an energy  $>$0.16 eV above the chemical potential couple to optical phonons, which in turn decay to acoustic phonons while the electronic system continuously re-thermalizes. The electron-to-optical-phonon cooling component gives rise to sub-picosecond cooling. Due to the hot-phonon bottleneck governed by the anharmonic coupling of optical to acoustic phonons, an approximately bi-exponential cooling occurs, where the second decay component has a characteristic timescale of a few picoseconds. The overall decay becomes slower for increasing initial electron temperature (higher incident fluence) due to the hot-phonon bottleneck. Our analytical model suggests that this mechanism will quickly become less efficient upon decreasing the ambient temperature $T_{\rm L}$, provided that also the amount of heating is small $\Delta T_{\rm e} < T_{\rm L}$. Thus, operating graphene with low incident fluence and at reduced ambient temperatures is likely a promising approach to optimize optoelectronic applications exploiting hot carriers in graphene. 
\\
\section{Methods}\label{methods}
	\paragraph{High-sensitivity transient absorption microscopy}
	The transient absorption microscope for measurements in the NIR is custom-built starting from a Er-doped fiber laser (Toptica-Femto fiber pro) generating 300~mW, 150~fs pulses centered at 1550~nm with 40~MHz repetition rate. 
	A portion of the output of the laser is used as the pump pulse and it is modulated with an acousto-optic modulator operating at 1~MHz. The NIR probe pulse is obtained by super continuum (SC) generation focusing part of the laser fundamental in a highly-nonlinear fiber. The high-energy component of the SC and the fundamental frequency are filtered out with a longpass filter cutting at 1600~nm, and the component at 1700 nm is selected with a double-grating monochromator with 5~nm spectral resolution. The pump and the probe are collinearly focused on the sample with an objective (Olympus-LCPLN-IR with magnification 100~X and NA = 0.85) over a spot size of about 1~$\mu$m. 
	The probe transmitted by the sample is collected with an achromatic doublet with a 8~mm focal length and detected by an InGaAs balanced amplified photodiode with 4~MHz bandwidth. The component of the probe at the modulation frequency is measured with a lock-in amplifier (HFLI, Zurich Instruments) using 300~ms effective time constant resulting in a $\Delta$ T/T sensitivity below 10$^{-6}$. Transient transmission dynamics is monitored by changing the pump-probe time delay with an optical delay line, while the pump-probe maps (images) at fixed time-delay are acquired by moving the sample with a motorized three-axis piezo-stack linear stage (Newport NPXYZ100). Image size of $120 \times 120$ pixels is used in the experiment. The width cross-correlation between pump and probe pulses at the sample is $\sim$~210 fs.
	
	The transient absorption measurements in the VIS on suspended graphene are performed using 400 nm pump and 800 nm probe pulses with 150 fs pulse duration at 76 MHz repetition rate. Both pulses are carefully overlapped and focused at the same focal plane with a 40x/0.6 NA objective to focal spot sizes of 0.6 and 0.9 $\mu$m, respectively. The pump beam is modulated with an optical chopper at 6.4 kHz. The probe is delayed temporally with a mechanical delay line and detected in transmission on a balanced photodiode via lock-in detection. Further details of the setup are described in Ref.~\citenum{block2019}.
	
	\paragraph{Raman spectroscopy}
	For Raman characterization, we used an inVia confocal Raman spectrometer from Renishaw plc, equipped with 473 nm and 532 nm CW laser sources. The laser beam was focused onto the sample through a 100x objective lens, with 0.89 NA. The nominal FWHM of the Gaussian beam at the focus is estimated to be $\sim$1 $\mu$m and the step size for the maps is set to 4 $\mu$m. The laser power hitting the sample in the selected configurations for the measurements were 1.6 mW for the 473 nm laser and 0.7 mW for the 532 nm. With an exposure time per pixel of 60 s, the fluence was 122 mJ/$\mu^2$ and 53 mJ/$\mu^2$, respectively. Depending on the measurement, a 1800 lines/mm grating or a denser 2400 lines/mm were used. The spectra are calibrated with respect to the Si peak at 520 cm$^{-1}$.

\section{Acknowledgments}
	The authors acknowledge funding from the European Union Horizon 2020 Programme under Grant Agreement No. 881603 Graphene Core 3.
	ICN2 was supported by the Severo Ochoa program from Spanish MINECO (Grant No. SEV-2017-0706). K.J.T. acknowledges funding from the European Union's Horizon 2020 research and innovation program under Grant Agreement No. 804349 (ERC StG CUHL), RyC fellowship No. RYC-2017-22330, IAE project PID2019-111673GB-I00, and financial support through the MAINZ Visiting Professorship.
	X.J. acknowledges the support from the Max Planck Graduate Center with the Johannes Gutenberg-Universit\"at Mainz (MPGC). J.Z. acknowledges the support from National Natural Science Foundation of China (No. 52072042). Z.L. acknowledges the support from National Natural Science Foundation of China (No. 51520105003). 		T.S. acknowledges support from the University of Liege under Special Funds for Research, IPD-STEMA Programme.
	M.J.V. gratefully acknowledges funding from the Belgian Fonds National de la Recherche Scientifique (FNRS) under PDR grant T.0103.19-ALPS. 
	Computational resources were provided by CECI (FRS-FNRS G.A. 2.5020.11); the Zenobe Tier-1 supercomputer (Gouvernement Wallon G.A. 1117545); and by a PRACE-3IP DECI grants 2DSpin and Pylight on Beskow (G.A. 653838 of H2020).

\section{Supporting Information}
	The Supporting Information is available free of charge on the
	ACS Publications website at DOI: .
	The hyperbolic cooling model, cooling dynamics probed with terahertz pulses, topography of encapsulated graphene, electron mobility of WSe$_2$-encapsulated graphene, Raman characterization of WSe$_2$-encapsulated graphene, differential reflectance of WSe$_2$-encapsulated graphene, cooling via disorder-assisted acoustic phonon scattering, cooling via optical phonons (PDF).

\newpage\hbox{}\thispagestyle{empty}\newpage
\onecolumngrid
\appendix

\section{Supporting Information}
\renewcommand{\thefigure}{S\arabic{figure}}
\setcounter{figure}{0}
\subsubsection{1. Hyperbolic cooling model}
The permittivities and the permittivity products of bulk hBN and WSe$_2$ have been calculated through density functional perturbation theory, as implemented in the \textit{abinit} code\cite{gonze2002first}, and the computational setup of Ref.~\cite{pike2018}. Along the trigonal axis zz, and, in-plane directions xx and yy, two bands of hyperbolicity can be observed in the low energy limit (energy less than 200 meV) due to hyperbolic optical phonons. The energy bands of hyperbolicity are strongly reduced in WSe$_2$ as compared to hBN, see Fig. S1b. 

\begin{figure*} [ht!p]
	\includegraphics[width=0.7\textwidth]{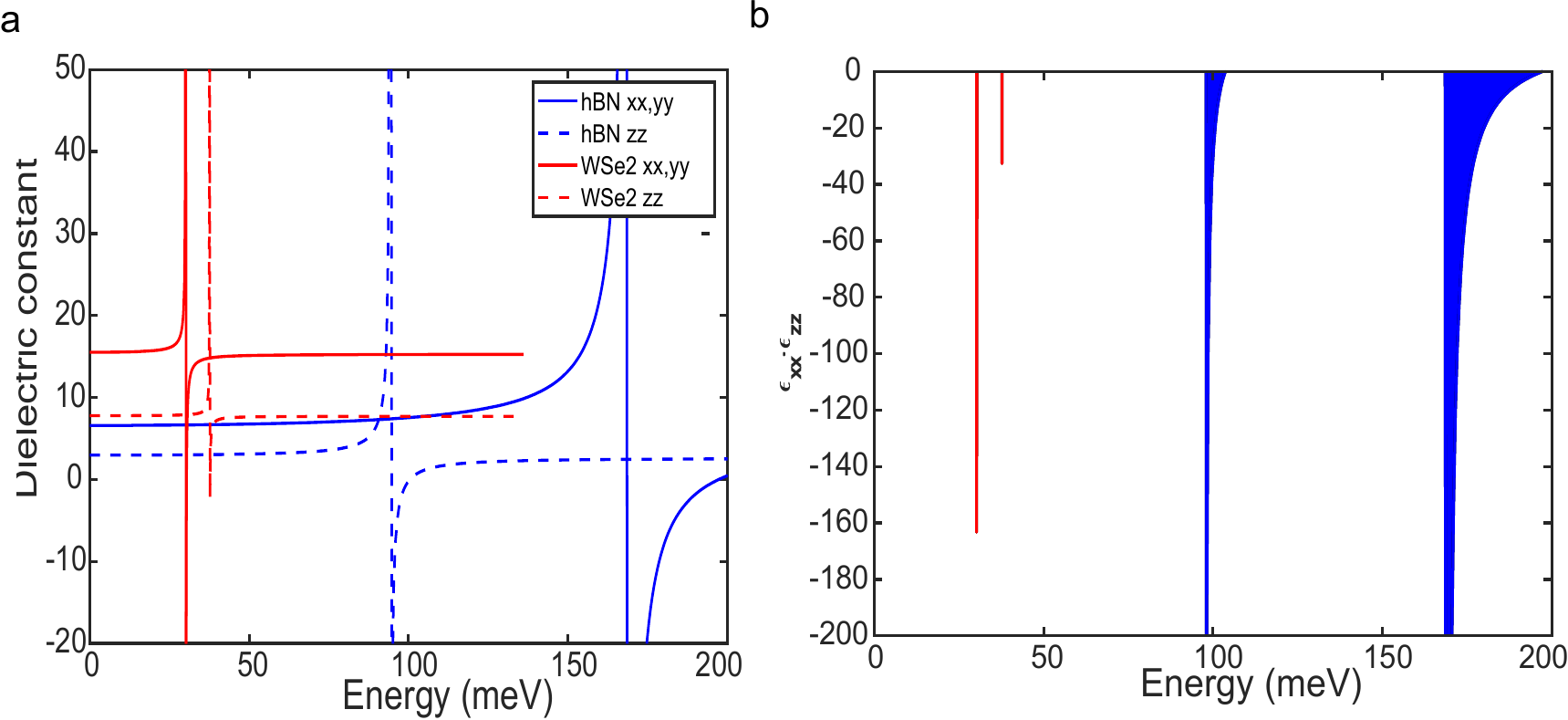}
	\caption{\textbf{Hyperbolic phonons in hBN and WSe$_2$.} 
		\textbf{a)} Calculated dielectric constants of hBN (blue lines) and WSe$_2$ (red lines) along the in plane directions (xx and yy, solid lines) and along the c-axis (zz, dashed lines). \textbf{b)} Hyperbolic bands in hBN (blue curves and areas) and WSe$_2$ (red lines and areas) corresponding to energy ranges of panel a in which the dielectric constants along the in-plane ($\epsilon_{xx}$) and out-of-plane ($\epsilon_{zz}$) directions have opposite sign.}
	\label{f:S0}
\end{figure*}

\begin{figure} [h!!!!!]
\centerline{\includegraphics[width=0.4\textwidth]{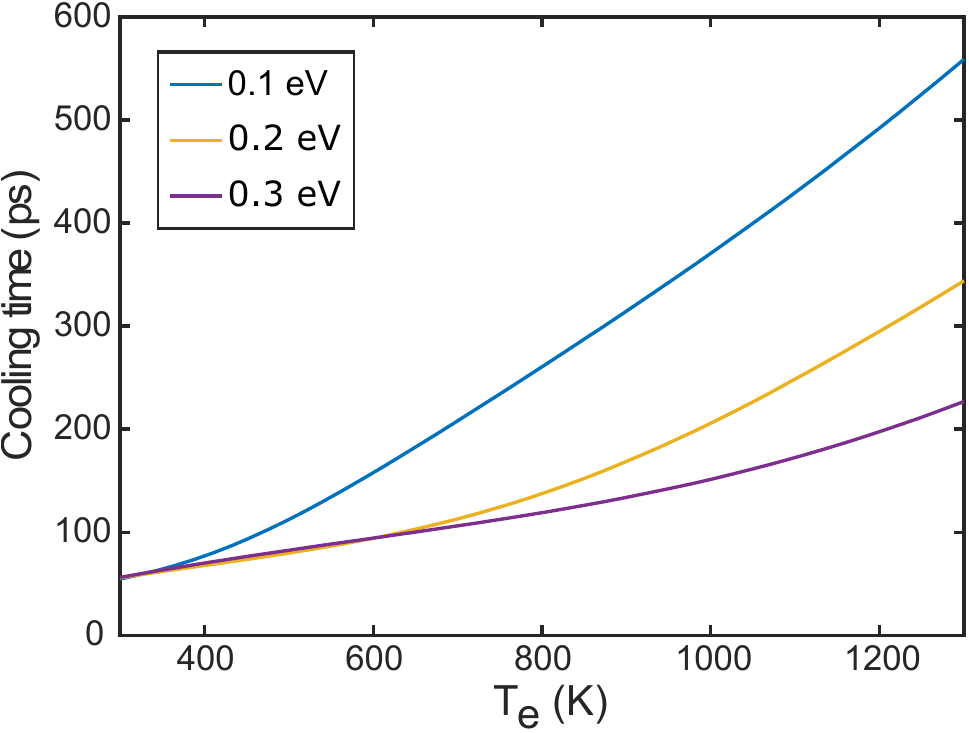}}
\caption{\textbf{Cooling time via near-field coupling with hyperbolic phonons of WSe$_2$.} 
Calculated hot electron cooling time considering the transfer of their excess-energy to the hyperbolic phonons of WSe$_2$ as a function of the initial electron temperature T$_e$, for three relevant values of the equilibrium chemical potential: 0.1, 0.2, 0.3 eV.}
\label{f:S1}
\end{figure}

The electron cooling time calculated from the dielectric constants dispersion in Fig. S1 using the hyperbolic cooling theory\cite{Principi2017,Tielrooij2018,Yang2018} is reported in Fig. \ref{f:S1}. The theory only considers the near-field coupling of the hot charge carriers of graphene with the hyperbolic phonons of WSe$_2$ as relaxation mechanism. Due to the reduced hyperbolicity, the cooling to hyperbolic phonons, which dominates in high-quality hBN-encapsulated single-layer graphene (SLG), is much less efficient in WSe$_2$-encapsulated SLG and the relaxation via emission of optical phonons eventually becomes the dominant process driving the cooling dynamics, in close analogy to suspended SLG. The expected cooling time via solely decay into hyperbolic phonons far exceeds, indeed, the few picoseconds recovery time measured by transient transmission in WSe$_2$-encapsulated graphene.

\newpage
\subsubsection{2. Cooling dynamics probed with terahertz pulses}

\begin{figure} [h!!!!!]
\centerline{\includegraphics[scale=1]{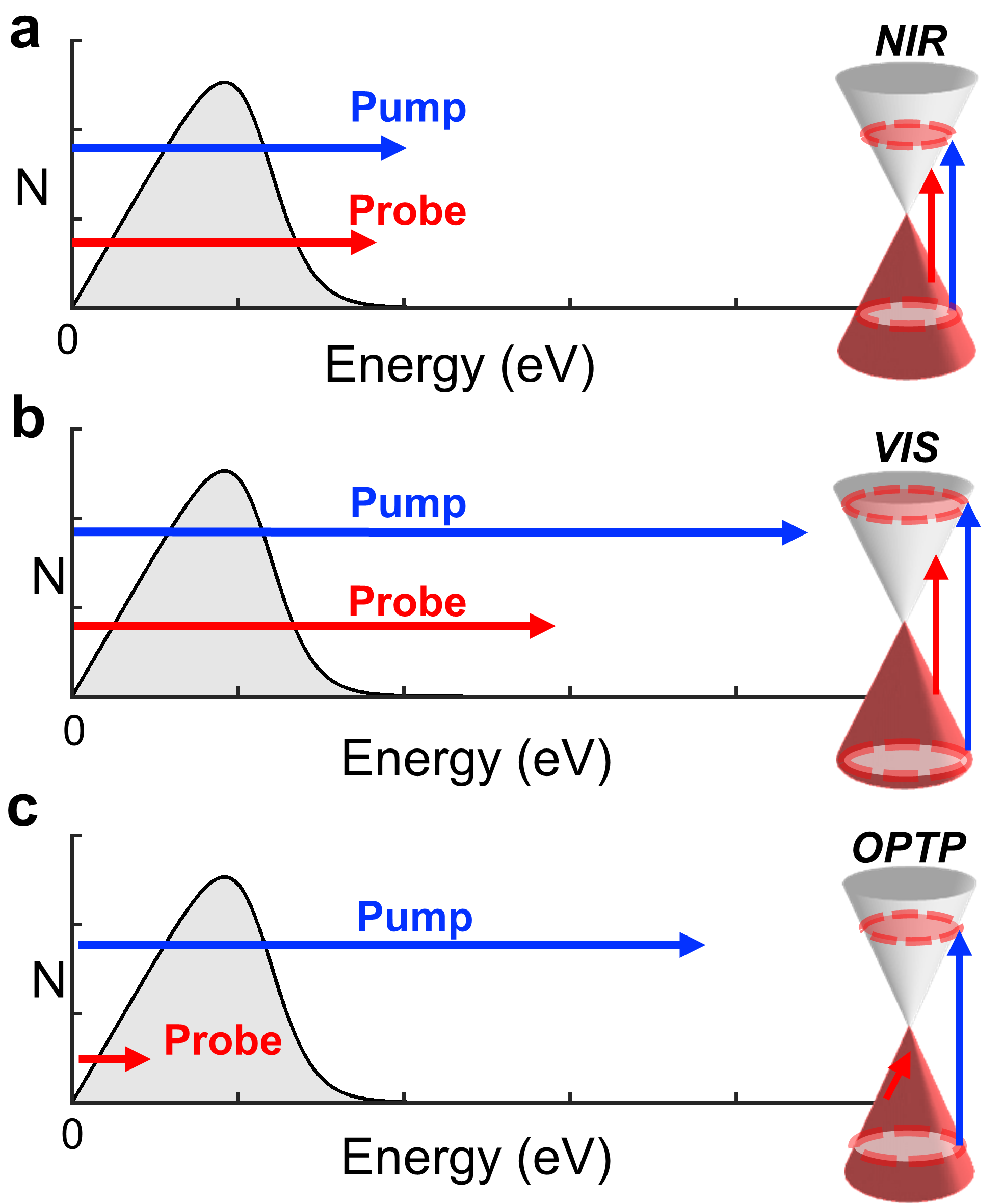}}
\caption{\textbf{Schematic illustration of the carrier density as a function of carrier energy compared to the pump and probe photon energies of the three experimental configurations that we used: (a)} near-infrared transient absorption (NIR), \textbf{(b)} visible transient absorption (VIS) and \textbf{(c)} optical-pump terahertz probe (OPTP). In all cases, the pump pulse is absorbed via interband transitions. For both TA configurations, the probe pulse is also associated with interband transitions, whereas the THz probe pulses are associated with the intraband conductivity. This is also clear in the accompanying Dirac cones with energy \textit{vs.}\ momentum.}
\label{FigS_PP}
\end{figure}

The cooling dynamics of suspended and WSe$_2$-encapsulated graphene is measured by optical pump-THz probe (OPTP) ultrafast spectroscopy, to monitor the dynamics of intraband absorption following the photo-excitation, as sketched in Fig. \ref{FigS_PP}. The spot size of the OPTP measurements is $\sim$ 1 mm. Accordingly, for suspended graphene, we could use the same sample as used for TA measurements, with the probe illuminating both suspended and supported graphene, considering that the signal will be dominated by the suspended graphene. 

\begin{figure} [h!!!!!]
\centerline{\includegraphics[width=0.4\textwidth]{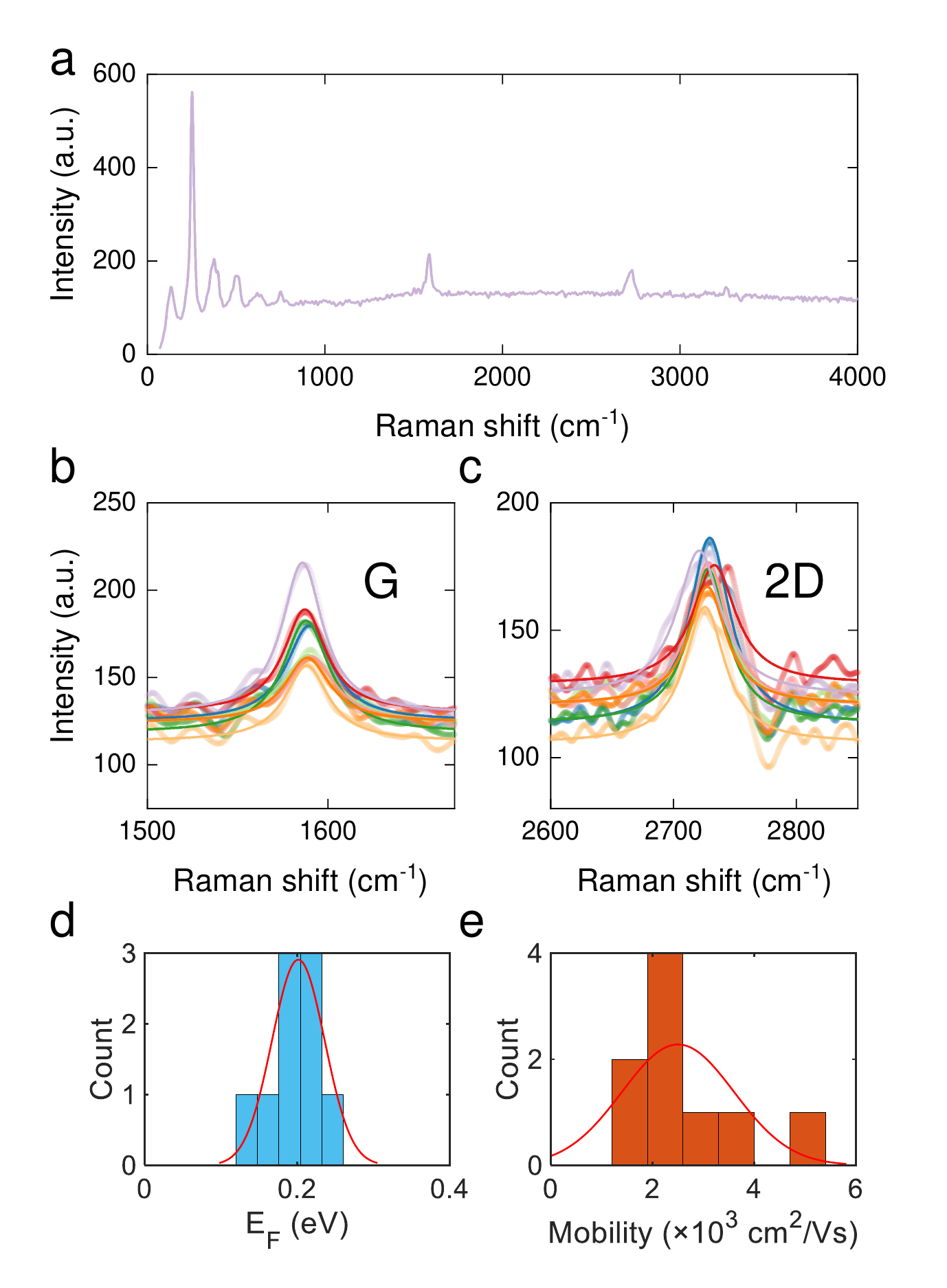}}
\caption{\textbf{Raman characterization of WSe$_2$-encapsulated graphene for THz experiments.} \textbf{ a)} Raman spectrum of the large area CVD grown graphene encapsulated by single layer WSe$_2$ displaying the low frequency fingerprints of WSe$_2$ and the G and 2D peaks from graphene, measured with excitation source at 488 nm.
\textbf{b-c)} G and 2D peaks of encapsulated graphene in different sample positions together with the best fit with Voigt functions to evaluate the G peak position $\omega_G$ and the 2D peak width $\Gamma_{2D}$. \textbf{c)} Doping concentration and mobility statistics extracted from $\omega_G$ and $\Gamma_{2D}$ of the peaks in panels b-c.}
\label{f:S7}
\end{figure}

For WSe$_2$-encapsulated graphene, we used a large-area CVD-grown sample, which was first characterized by Raman spectroscopy for estimating the mobility and initial doping concentration, see Fig. \ref{f:S7}.

From the average G peak position $\omega_{G}$ = 1588.46 cm$^{-1}$, we extract a Fermi energy  $|E_{\rm F}|$ of 0.2 eV, while from the average width of the 2D peak, $\Gamma_{2D}$ = 39.59 cm$^{-1}$, we infer a mobility of $\sim$ 2300 cm$^2$V$^{-1}$s$^{-1}$ \cite{Robinson2009}.

\begin{figure*} [h!!!!!]
\centerline{\includegraphics[width=0.9\textwidth]{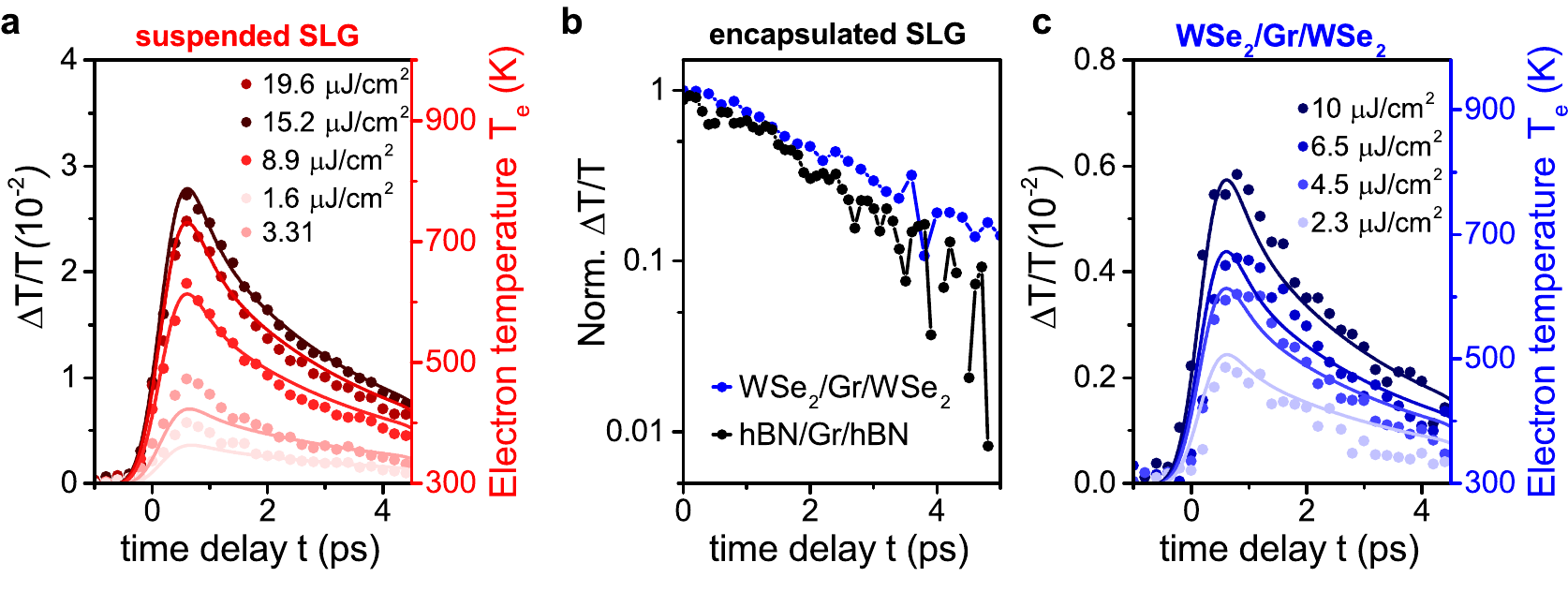}}
\caption{ \textbf{Cooling dynamics of suspended and WSe$_2$-encapsulated graphene using THz pulses.}
\textbf{a} OPTP dynamics for high-quality suspended graphene with pump photon energy of 1.55 eV (800 nm) and probe photon energy around 4 meV (300 $\mu$m) together with the calculated cooling dynamics for the electron temperature $T_{\rm e}$ (solid lines).
\textbf{b)} Normalized $\Delta T/T$ dynamics measured through OPTP spectroscopy, with pump photon energy of 1.55 eV (800 nm) and probe photon energy around 4 meV (300 $\mu$m). We compare large-area graphene encapsulated by monolayer WSe$_2$ (blue dots) and by hBN (black dots), finding faster cooling in the case of hBN encapsulation, due to out-of-plane cooling \cite{Principi2017,Tielrooij2018,Yang2018}. 
\textbf{c} OPTP dynamics for large-area WSe$_2$-encapsulated graphene together with the calculated cooling dynamics for the electron temperature $T_{\rm e}$ (solid lines).}
\label{Fig6}
\end{figure*}

In Fig. \ref{Fig6} a we show the OPTP dynamics for suspended graphene for a range of fluences. We note that we see much less of the fast decay component that is rather prominent in the TA measurements. The reason for this is twofold. First of all, the superlinear relation between TA signal and $\Delta T_{\rm e}$ makes the TA signal initially decay faster than the decay of the carrier temperature. Secondly, the time resolution of the OPTP measurement is $\sim$300 fs, compared to $\sim$100 fs for TA measurements, which broadens the initial fast decay. 

In Figs. \ref{Fig6} b-c, we show the comparison of the relaxation dynamics in graphene encapsulated by WSe$_2$ and by hBN. In agreement with the NIR TA results (see Fig.\ 2b), we observe faster cooling for hBN encapsulation due to out-of-plane cooling \cite{Principi2017,Tielrooij2018,Yang2018}. With THz probe pulses, the transmission increases (for $E_{\rm F} >$0.1 eV), as observed and explained in Refs.\cite{Tielrooij2013,Frenzel2014,Tomadin2018} and the change in transmission scales roughly linearly with the carrier temperature change $\Delta T_{\rm e}$, in contrast with the superlinear relation for TA measurements. 

We compare the OPTP relaxation dynamics with the calculations based on the cooling mechanism involving high-energy ($>$0.16 eV) electrons cooling to optical phonons, which in turn cool to acoustic phonons, while the carrier system continuously re-thermalizes. We describe these dynamics with the same model as before, using $E_{\rm F}$  = 0.15 eV and an  optical phonon lifetime of 1.2 ps. Here, we have assumed a perfectly linear relation between OPTP signal and $\Delta T_{\rm e}$. We note that there is some underestimation of the signal magnitude for low fluences, and some overestimation for higher fluences, which shows that the assumption of linear scaling between signal and temperature change is a simplification, as it likely weakly sub-linear. However, the reasonable agreement  between calculated and measured dynamics, confirms the validity of this intrinsic cooling mechanism. The slightly shorter optical phonon lifetime we extract by comparing the experimental data with our cooling model, is likely the result of the assumption of linear scaling between OPTP signal and $\Delta T_{\rm e}$

Despite the agreement with the model based on the scattering with optical phonons and on the electron re-thermalization, the contribution of supercollision mechanism to the carrier relaxation dynamics cannot be completely ruled out  in the case of the CVD grown WSe$_2$ encapsulated graphene because the carrier mobility is $< 10,000$ cm$^2 V^{-1}s^{-1}$. 

\subsubsection{3. Topography of encapsulated graphene}

\begin{figure} [ht!p]
	\centerline{\includegraphics [width=0.5\textwidth]{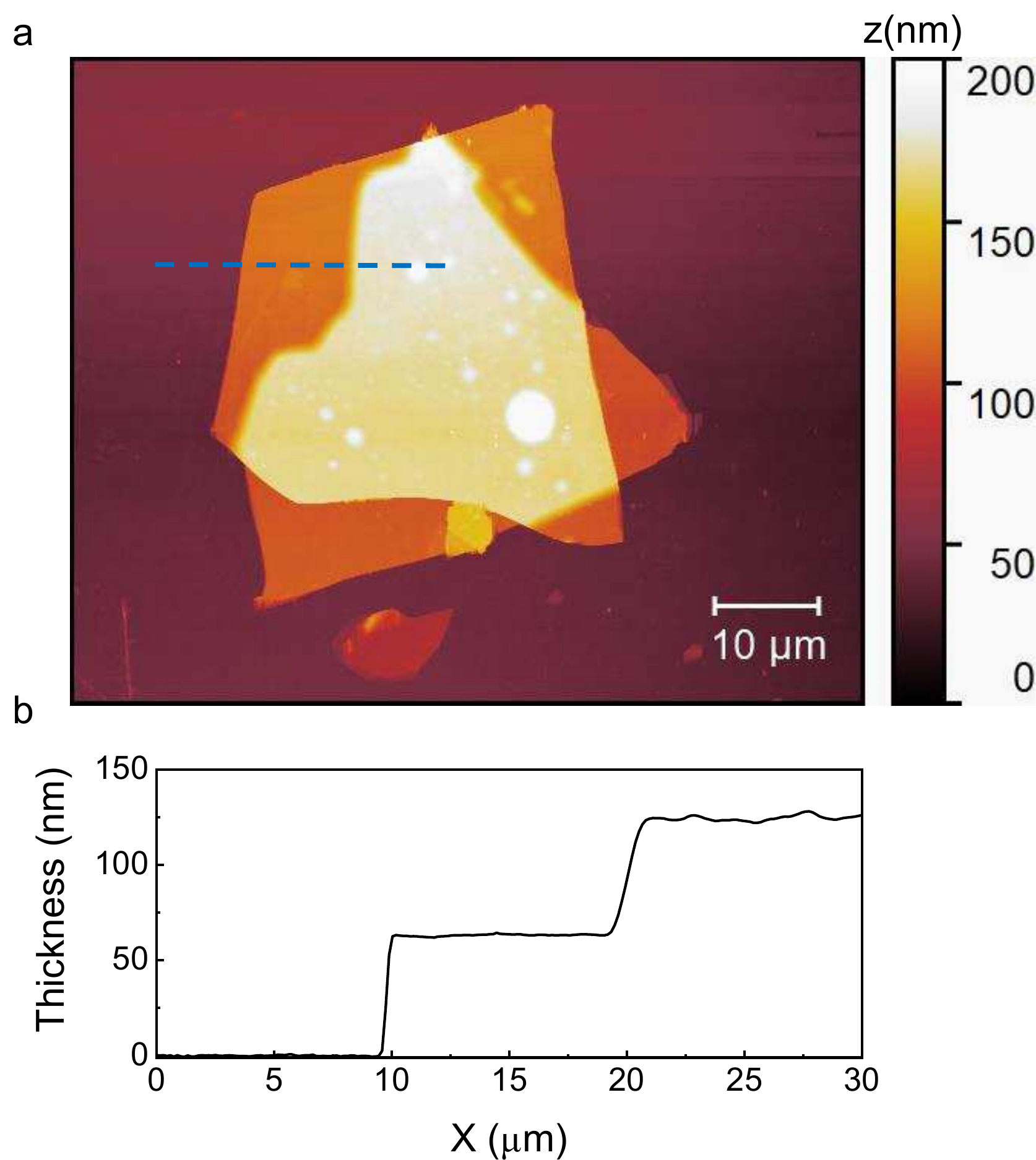}}
	\caption{\textbf{Topography of WSe$_2$ encapsulated SLG.} 
		\textbf{a)} AFM image of encapsulated SLG corresponding to Fig.1 of the main text in a different color scale. \textbf{b)} Height profile extracted from the topography in panel a along the blue dashed line assuming the substrate level as zero.}
	\label{f:S2}
\end{figure}

The thickness of the WSe$_2$ encapsulating flakes is determined from the AFM image in Fig. \ref{f:S2} by analysing a line profile cut from the map, which includes the substrate, whose height is assumed as z= 0, and both the bottom and top WSe$_2$ layers. The bottom and top WSe$_2$ layers have a thickness of 63 nm and 61 nm respectively, as obtained  from the total thickness of the heterostructure z= 124 nm. 

\newpage
\subsubsection{4. Electron mobility of WSe$_2$-encapsulated graphene}
\begin{figure} [h!!!!!]
\centerline{\includegraphics[width=0.5\textwidth]{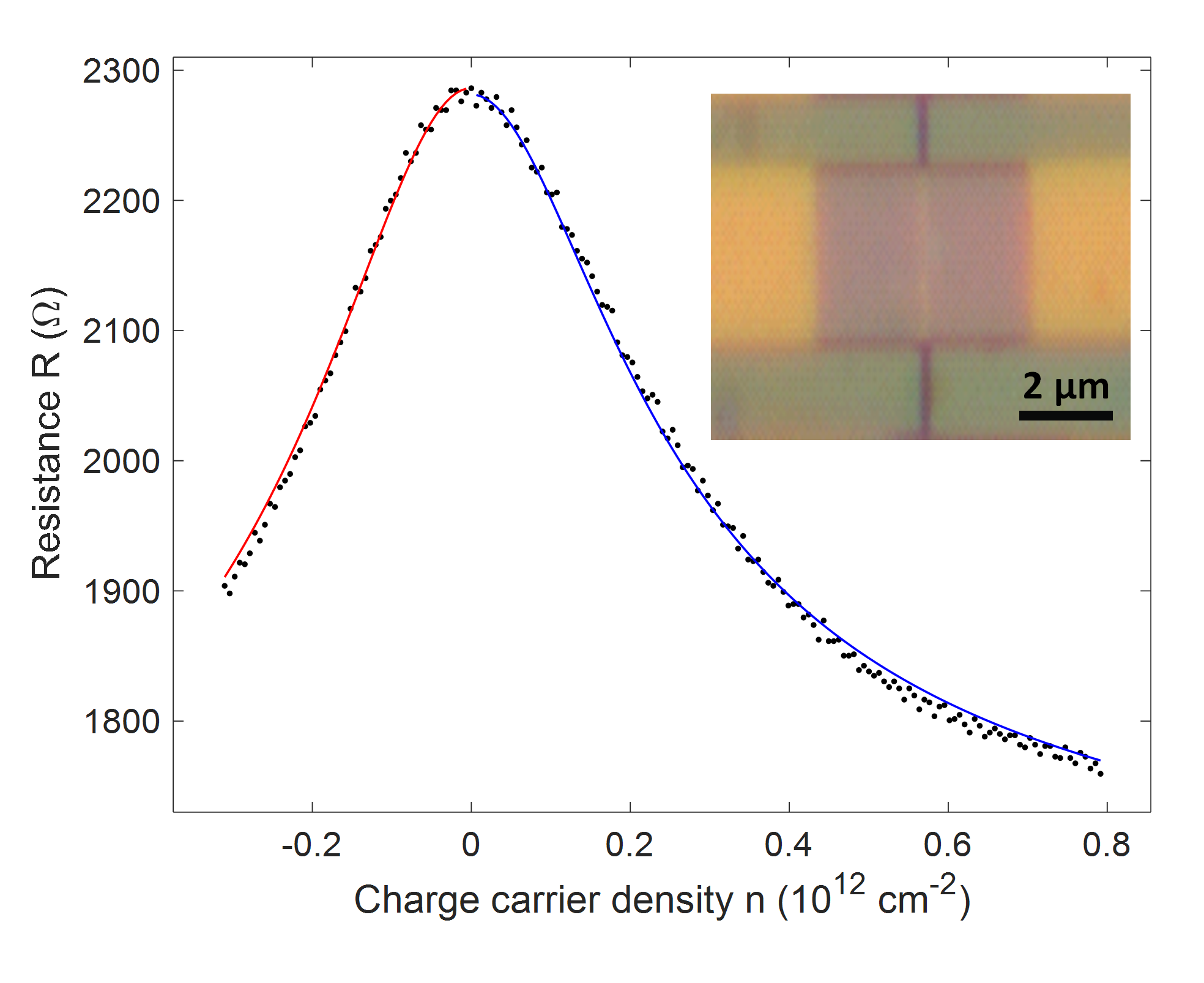}}
\caption{\textbf{Mobility of WSe$_2$ encapsulated graphene.} Charge transfer curve (black dots) for WSe$_2$ encapsulated graphene measured at room temperature applying 1 mV voltage bias to source-drain contacts of a dual-gate field effect transistor together with the best fit function fr p-doping corresponding to hole mobility $\mu_h$= 36,000 cm$^2$V$^{-1}$s$^{-1}$ (red solid line), and, n-doping giving electron mobility $\mu_e$= 39,000 cm$^2$V$^{-1}$s$^{-1}$ (blue solid line). See inset for AFM image of the device.}
\label{f:S5}
\end{figure}

The mobility of WSe$_2$ encapsulated graphene is determined by transport measurements reported in Fig. \ref{f:S5}. A dual-gate field effect transistor based on encapsulated graphene is fabricated to monitor the resistance as a function of the charge carrier density $n_e$. The transistor channel has length L= 4.5 $\mu$m and width W= 4 $\mu$m. The top WSe$_2$ is 12 nm thick while the bottom layer is 34 nm. Efficient gating is achieved by stacking the heterostructure on top of an additional layer of hBN of thickness 25 nm. The resistance of graphene shows the typical ambipolar conduction with electron mobility $\mu_e\approx$ 39,000 cm$^2$V$^{-1}$s$^{-1}$, and hole mobility $\mu_h\approx$ 36,000 cm$^2$V$^{-1}$s$^{-1}$.

\newpage
\subsubsection{5. Raman characterization of WSe$_2$-encapsulated graphene }
\begin{figure} [h!!!!!]
\centerline{\includegraphics[width=0.68\textwidth]{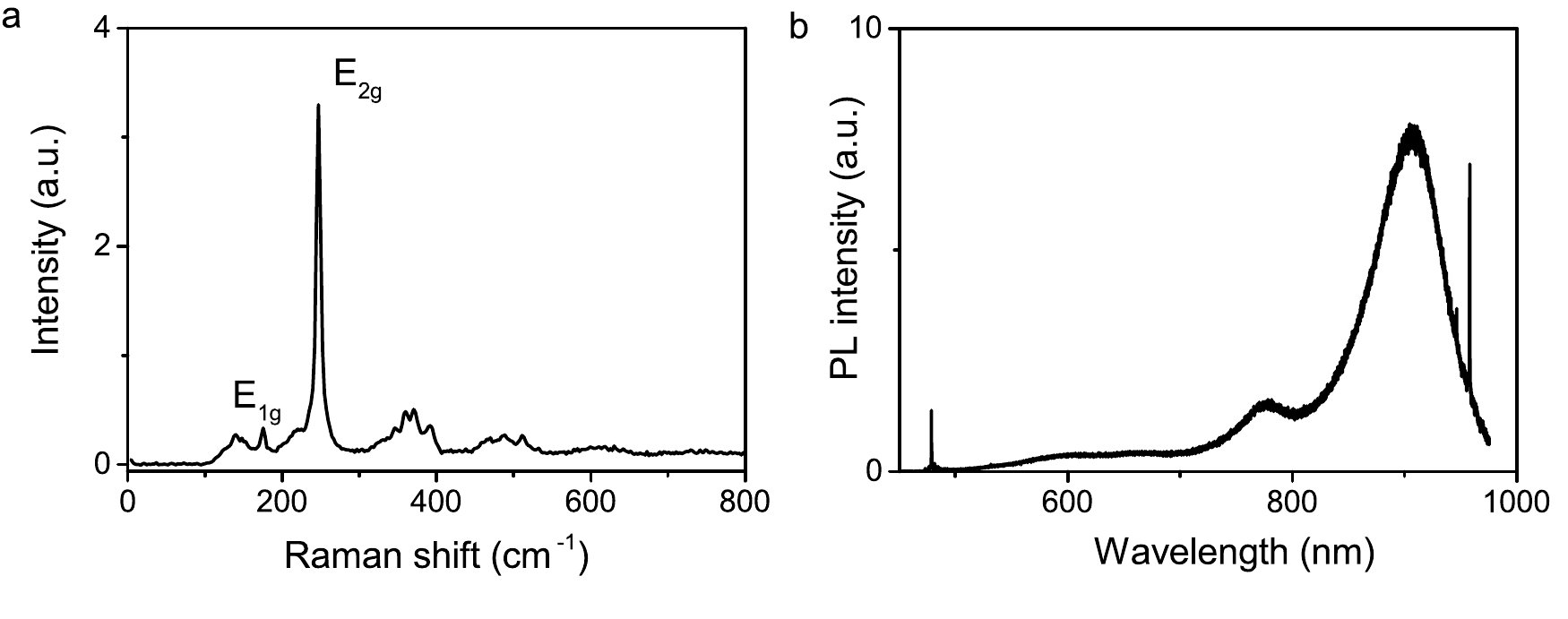}}
\caption{\textbf{Raman characterization of encapsulated SLG} \textbf{a)} Raman spectroscopy measurements of the encapsulated SLG revealing the low-energy vibrational fingerprints of bulk WSe$_2$\cite{Terrones2014}. \textbf{b)} Photoluminescence (PL) of encapsulated SLG under 473 nm laser photoexcitation.}
\label{f:S3}
\end{figure}

The encapsulated SLG exhibits the vibrational fingerprints of bulk WSe$_2$ in the low frequency range, see Fig. \ref{f:S3}a, corresponding to E$_{1g}$ and $E_{2g}$ modes\cite{Terrones2014}.
The encapsulated SLG under 473 nm laser excitation exhibits photoluminescence signal in Fig. \ref{f:S3}b with a peak at 776 nm due to the radiative recombination of the A exciton of bulk WSe$_2$ and one more intense peak at 908 nm attributable to the CaF$_2$ substrate.

\begin{figure} [h!!!!!]
\centerline{\includegraphics[width=0.3\textwidth]{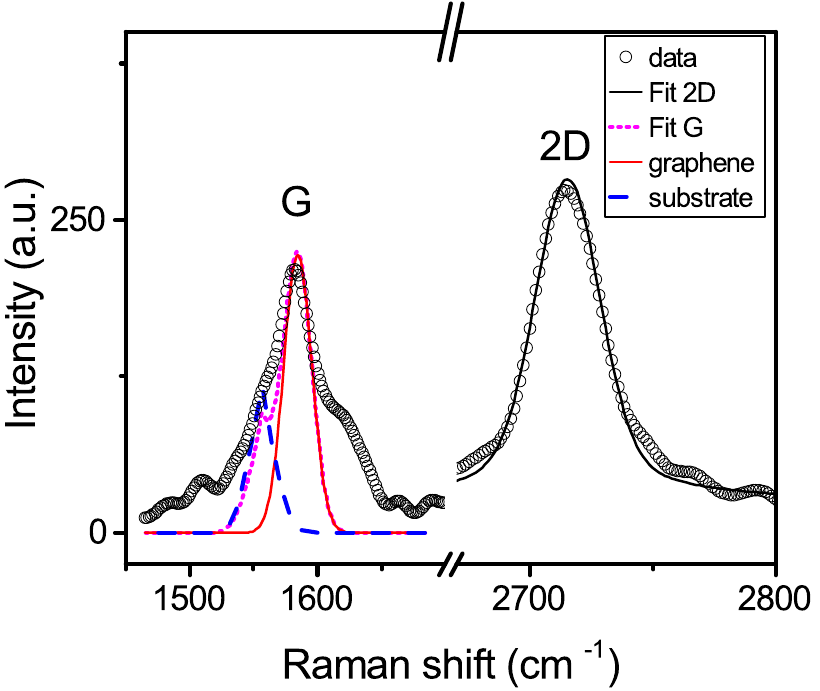}}
\caption{\textbf{Raman spectrum of WSe$_2$-encapsulated graphene} Raman spectrum of WSe$_2$/Gr/WSe$_2$ acquired with 473 nm laser source (black dots). The 2D peaks is analysed with PseudoVoigt functions centered at $\omega_{2D}$ with width $\Gamma_{2D}$ (black solid line). The peak at around 1550 $cm^{-1}$ is analysed as the sum (magenta dotted line) of two peaks: one from the substrate, centered at about 1556 cm$^{-1}$ (blue dashed line) and the G-peak of graphene (red line).}
\label{f:S4}
\end{figure}

Despite the intense luminescence background, the vibrational fingerprints of graphene can be identified in the encapsulated graphene. 
One exemplary spectrum acquired on fully encapsulated graphene with a 473 nm laser is reported in Fig. \ref{f:S4}, showing the 2D and G peaks. The 2D peak is fit with pseudoVoigt function centered at $\omega_{2D}$= 2714.5 cm$^{-1}$ with width $\Gamma_{2D}$= 25.7 cm$^{-1}$.
%for the spectrum acquired at the first position (point 1), and, at $\omega_{2D}$= 2715.2 cm$^{-1}$ with $\Gamma_{2D}$= 30.9 cm$^{-1}$ for the second position (point 2). 
The substrate exhibits a Raman peak at around 1550 cm$^{-1}$, which has maximum amplitude outside the heterostructure. Accordingly, to retrieve the position and the width of the G peak, we analyse the peak at 1550 cm$^{-1}$ as the sum of the peak from the substrate and of the G-peak, both described by pseudoVoigt functions. The retrieved energy position and width of the G peak are equal to $\omega_G$= 1585.2 cm$^{-1}$ and $\Gamma_G$= 24.2 cm$^{-1}$.
% for point 1, and to $\omega_G$= 1584.6 cm$^{-1}$ and $\Gamma_G$= 24.2 cm$^{-1}$ for point 2.

\newpage
\subsubsection{6. Differential reflectance of WSe$_2$-encapsulated graphene}
\begin{figure} [h!!!!!]
\centerline{\includegraphics[width=0.5\textwidth]{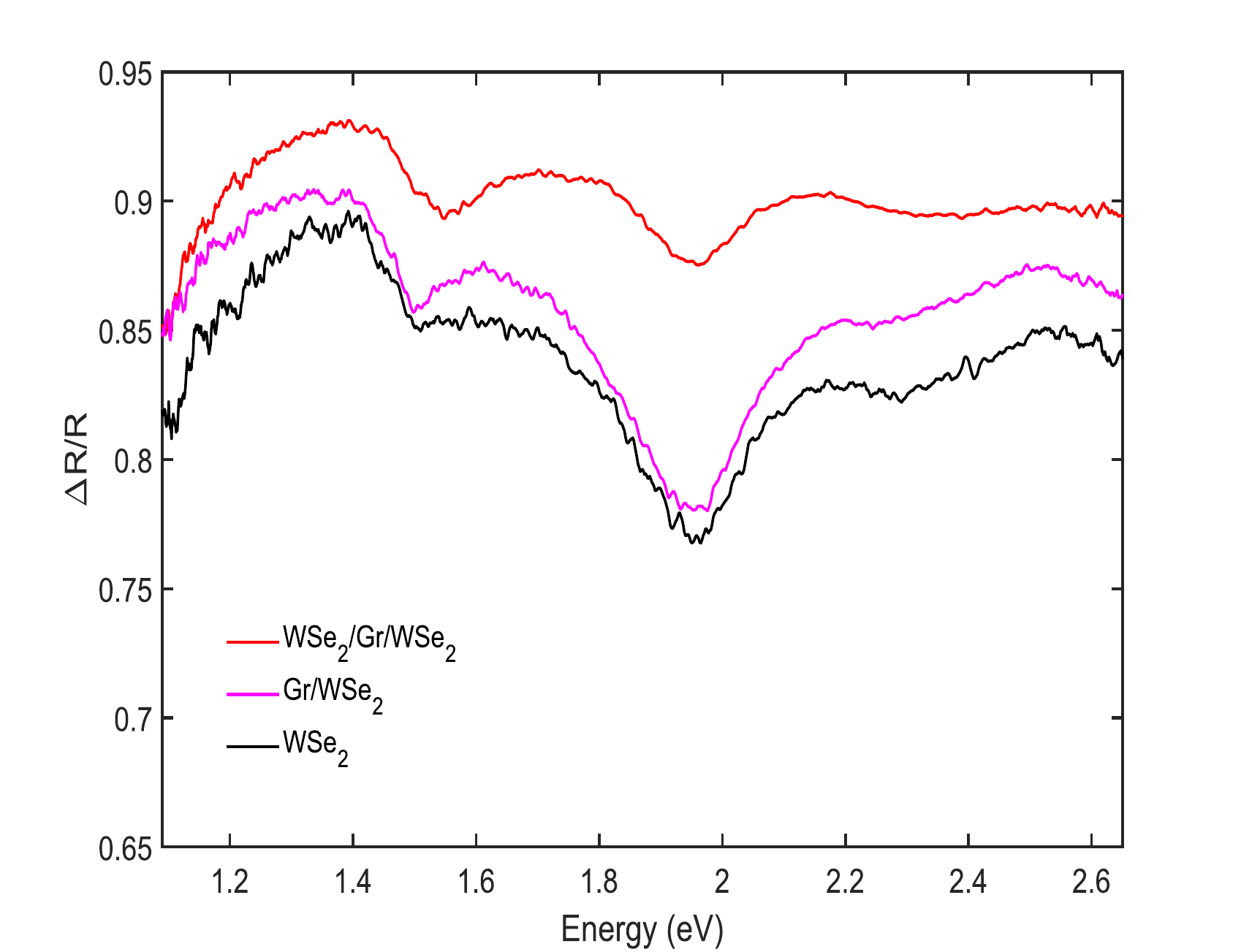}}
\caption{\textbf{Static reflectance of WSe$_2$-encapsulated graphene.}
The static reflectance $\Delta R/R={\frac{R-R_0}{R_0}}$ is evaluated by comparing the reflectance of the substrate R$_0$ with the reflectance $R$ of the heterostructure. We compare the response at the WSe$_2$ bottom layer (black line), with that of fully WSe$_2$-encapsulated graphene (WSe$_2$/Gr/WSe$_2$, red-line) and with half-encapsulated graphene (Gr/WSe$_2$, magenta line).}
\label{f:S6}
\end{figure}

The reflectance of the WSe$_2$ encapsulated graphene in the visible and near-infrared range is reported in Fig. \ref{f:S6}. The absorption dips for photon energy 1.9 and 1.5 eV, are due to the excitations of bulk excitons of WSe$_2$: A at lower and B at higher energy. 
The reflectance in the near-infrared region, far from the tail of the A-exciton, is completely attributable to graphene such that no difference is observed between the half (Gr/WSe$_2$) and the full (WSe$_2$/Gr/WSe$_2$) encapsulated sample. The pump and probe photons used in the transient transmission experiment are tuned at even lower photon energies, where only graphene is supposed to absorb.

\subsubsection{7. Cooling via disorder-assisted acoustic phonon scattering}
We calculate the cooling time through disorder-assisted scattering with acoustic phonons following Ref. \cite{Graham2013}, which was based on Ref. \cite{Song2012a}. For this calculation, we assume that the electrical mobility is limited by disorder scattering, although there is actually a large contribution from long-range Coulomb scattering. This type of scattering does not contribute to disorder-assisted cooling, which means that by assuming a disorder-dominated electrical mobility we obtain a lower bound for the cooling time. We calculate the cooling time using $\tau_{\rm cool}^{\rm SC} = (3\frac{A}{\alpha} \cdot T_{\rm l})^{-1}$, where $T_{\rm l}$ is the lattice temperature (300 K), and $\frac{A}{\alpha} = \frac{2~ \lambda~k_{\rm B}}{3~k_{\rm F}~\ell~\hbar}$. Here, $k_{\rm B}$ is Boltzmann's constant, Fermi momentum $k_{\rm F} = \frac{E_{\rm F}}{\hbar~v_{\rm F}}$, with $v_{\rm F}$ the Fermi velocity, and the mean free path is $\ell = v_{\rm F}~\tau_{\rm ms}$, where the momentum scattering time is given by $\tau_{\rm ms} = \frac{\mu~E_{\rm F}}{e~v_{\rm F}^2}$. Finally, $\lambda = \frac{D^2~2~E_{\rm F}}{\rho~v_{\rm s}^2~\pi~(\hbar~v_{\rm F})^2}$, where $\rho$ is the mass density and $v_{\rm s}$ is the sound velocity. For the deformation potential $D$ we chose 15 eV. With these equations, we obtain a cooling time $\tau_{\rm cool}^{\rm SC}$ = 22 ps for WSe$_2$-encapsulated graphene with $\mu \approx$50,000 cm$^2$V$^{-1}$s$^{-1}$ and $E_{\rm F} \approx$0.1 eV. For suspended graphene with $\mu \approx$17,000 cm$^2$V$^{-1}$s$^{-1}$ and $E_{\rm F} \approx$0.18 eV, we find $\tau_{\rm cool}^{\rm SC}$ = 14 ps. Given these timescales, and the fact that they are lower bounds, we conclude that disorder-assisted cooling does not play an important role in our samples.  

\begin{figure} [h!!!!!]
\centerline{\includegraphics[width=0.5\textwidth]{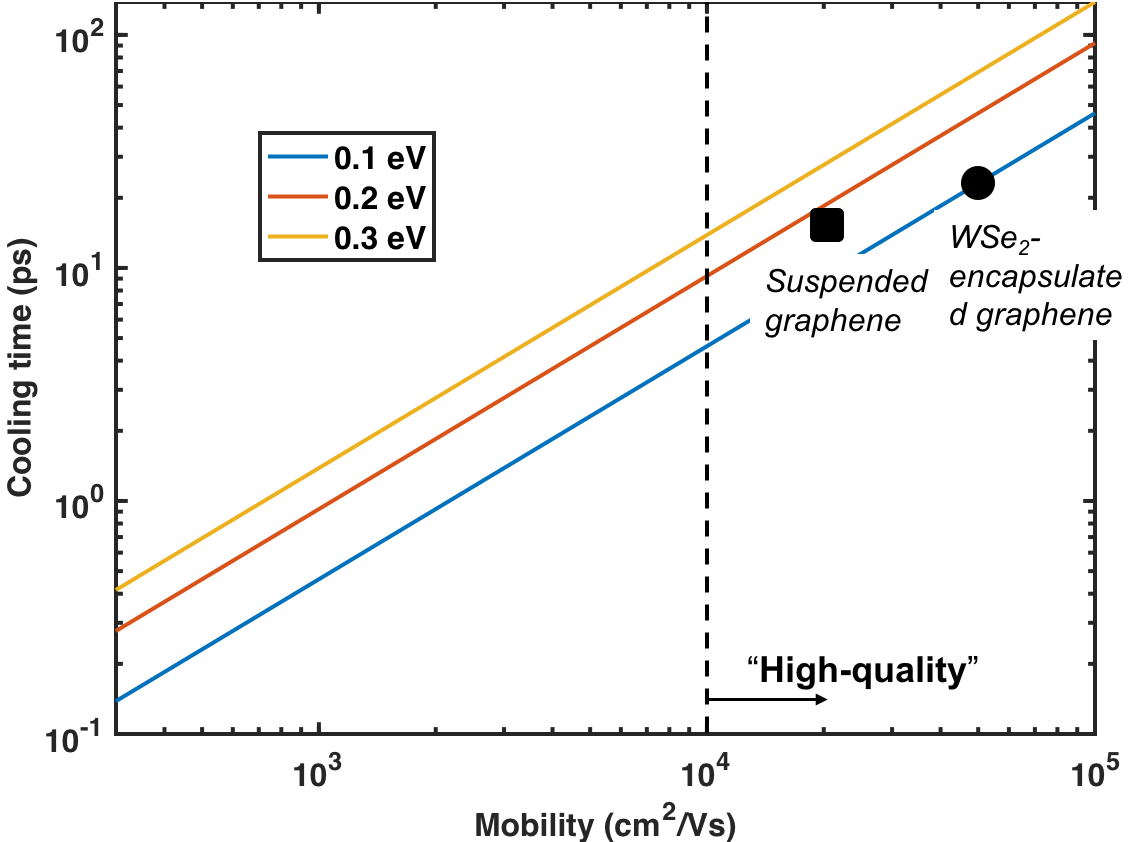}}
\caption{\textbf{Calculated ``supercollision'' cooling times}
}
\label{f:S11}
\end{figure}

\subsubsection{8. Cooling via optical phonons}

The cooling of electrons to optical phonons is described by means of the following rate equations for the electronic and optical-phonon temperatures, $T_e(t)$ and $T_\alpha(t)$, respectively:
\be
\left\{
\begin{array}{l}
	{\displaystyle C_e\big(T_e(t)\big) \partial_t T_e(t) = -\hbar \sum_{\alpha} \omega_\alpha {\cal R}_\alpha\big(T_e(t),T_\alpha(t)\big) }
	\vspace{0.2cm}\\
	{\displaystyle {\cal D}\big(\omega_\alpha,T_\alpha(t)\big) \partial_t T_\alpha(t) = \frac{{\cal R}_\alpha\big(T_e(t),T_\alpha(t)\big)}{M_\alpha\big(T_e(t)\big)} - \gamma_\alpha\big[n_\alpha\big(T_\alpha(t)\big)-n_\alpha(T_\alpha^{(0)})\big] }
\end{array}
\right.
~,
\ee
where $C_e(T_e)$ is the electron heat capacity, the sum is over phonon modes (labelled by $\alpha$), $\omega_\alpha$ is the frequency of mode $\alpha$, and ${\cal R}_\alpha(T_e,T_\alpha)$ is the rate of $\alpha$-phonon emission. Its expression is derived in the following section from a microscopic Boltzmann-equation approach [see Eq.~(\ref{eq:rate_phonon_emission_def})].
Furthermore, ${\cal D}(\omega_\alpha,T_\alpha) = \partial n_\alpha(T_\alpha)/(\partial T_\alpha)$, where $n_\alpha(T_\alpha) = [e^{\hbar \omega_\alpha/(k_{\rm B}T_\alpha)}-1]^{-1}$ is the phonon occupation function, while $T_\alpha^{(0)}$ is the equilibrium phonon temperature (in our calculations, the lattice temperature $T_L$) and $\gamma_\alpha$ is a phenomenological damping rate encoding, e.g., the decay of optical phonons into acoustic ones. In our calculation we consider phonon modes at the $\Gamma$ and $K/K'$ points of the phonon Brillouin zone, which scatter electrons in the same valley or between different valleys, respectively.
Finally ($v_{\rm F} \simeq 10^6~{\rm m/s}$ is the electron Fermi velocity),
\be \label{eq:phonon_density_def}
M_\alpha(T_e) = \frac{N_\alpha}{4\pi}\left[\left(\frac{\varepsilon_{\rm max}(T_e)}{\hbar v_{\rm F}}\right)^2 - \left(\frac{\varepsilon_{\rm min}(T_e)}{\hbar v_{\rm F}}\right)^2\right],
\ee
is the phonon density of states. $M_\alpha(T_e)$ measures the size of the portion (an annulus) of phonon Brillouin zone that is heated in the electron-cooling process and depends, assuming the phonon dispersion to be flat, on the maximum momentum that can be exchanged between electrons and phonons. As such, $M_\alpha(T_e)$ depends on the electron temperature. In Eq.~(\ref{eq:phonon_density_def}), $N_\alpha$ is the phonon degeneracy ($N_\alpha=2$ for both $\Gamma$- and $K$-phonons -- see below), while $\varepsilon_{\rm min}(T_e)$ and $\varepsilon_{\rm max}(T_e)$ are the minimum and maximum phonon energies, respectively. Their expressions are derived in the following.

The time-evolution of electronic temperature dynamics is converted into the dynamics of the differential conductance via~\cite{Stauber2008,Falkovsky2008}
\be
\frac{\Delta T}{T}(t) = - \frac{4\pi}{c} \frac{2}{n_1 + n_2} \Delta\sigma\big(\omega_{\rm ph},T_e(t)\big)
~,
\ee
where $n_1$ and $n_2$ are the refractive indices of the media above and below the graphene sheet, $\hbar\omega_{\rm ph}$ is the photon excitation energy, $\Delta\sigma(\omega_{\rm ph},T_e) = \sigma(\omega_{\rm ph},T_e) - \sigma(\omega_{\rm ph},T_L)$ and 
\be
\sigma(\omega_{\rm ph},T_e) = \frac{e^2}{4\hbar} \left[f(\frac{-\hbar \omega_{\rm ph}/2 -\mu}{k_{\rm B} T_e}) - f(\frac{\hbar \omega_{\rm ph}/2 -\mu}{k_{\rm B} T_e})\right]
~,
\ee
is the electronic optical conductivity. Here, $f(x) = (e^x+1)^{-1}$ if the Fermi-Dirac distribution function, while $\mu$ is the time-dependent electron chemical potential (see below).

\subsection{Microscopic derivation of the electron-phonon cooling rate}
We start from the Boltzmann equation satisfied by the electron distribution function, $f_{{\bm k},\lambda}$ (momentum ${\bm k}$, band $\lambda=\pm$), in a homogeneous system and in the absence of forces: 
\begin{eqnarray} \label{eq:Boltmann_def}
\partial_t f_{{\bm k},\lambda} = \sum_\alpha {\cal I}_{{\bm k},\lambda,\alpha}~.
\end{eqnarray}
The right-hand side of this equation is the sum of collision integrals accounting for the scattering between electrons and given phonon modes:
\begin{eqnarray}
{\cal I}_{{\bm k},\lambda,\alpha} = \sum_{{\bm k}',\lambda'} \big[ f_{{\bm k},\lambda} (1 - f_{{\bm k}',\lambda'}) W_{{\bm k},\lambda\to {\bm k}',\lambda'}^{(\alpha)} -  f_{{\bm k}',\lambda'} (1 - f_{{\bm k},\lambda}) W_{{\bm k}',\lambda'\to {\bm k},\lambda}^{(\alpha)} \big]
~,
\end{eqnarray}
where the transition probability, within a Fermi-golden rule approach, is
\begin{eqnarray}
W_{{\bm k},\lambda\to {\bm k}',\lambda'}^{(\alpha)} &=& \frac{2\pi}{\hbar V} \sum_{{\bm q},\nu} \big| U_{\lambda,\lambda',\alpha} ({\bm k}, {\bm k}',{\bm q}) \big|^2 \big[ 
(n_{{\bm q},\alpha} + 1) \delta(\varepsilon_{{\bm k},\lambda} - \varepsilon_{{\bm k}',\lambda'} -\hbar \omega_{{\bm q},\alpha}) \delta({\bm k} - {\bm k}' - {\bm q})
\nonumber\\
&+&
n_{{\bm q},\alpha} \delta(\varepsilon_{{\bm k},\lambda} - \varepsilon_{{\bm k}',\lambda'} + \hbar \omega_{{\bm q},\alpha}) \delta({\bm k} - {\bm k}' + {\bm q})
\big]
~.
\end{eqnarray}
Here $\varepsilon_{{\bm k},\lambda}$ is the electron band energy, $n_{{\bm q},\alpha}$ is the non-equilibrium photon distribution functions and $U_{\lambda,\lambda',\alpha} ({\bm k}, {\bm k}',{\bm q})$ is the electron-phonon interaction. There are two degenerate optical-phonon modes at the $\Gamma$ point, longitudinal and transverse, both with energy $\hbar \omega_\Gamma = 196~{\rm meV}$, which induce intra-valley electronic transitions. For these,~\cite{Sohier2014}
\be
\big|U_{\lambda,\lambda',\Gamma} ({\bm k}, {\bm k}',{\bm q})\big|^2 = g_{\Gamma}^2 \left[1 \pm \lambda \lambda' \cos\left(\varphi_{\bm k} + \varphi_{{\bm k}'} - 2\varphi_{\bm q}\right)\right]
~,
\ee
where the plus (minus) sign applies to longitudinal (transverse) phonons. We will assume that these modes are equally populated, {\it i.e.} their temperatures are equal to $T_\Gamma(t)$. In what follows, we will add up their contributions to the heat dissipated into a singe function ${\cal Q}_\Gamma(T_e,T_\Gamma)$.

Similarly, the modes at the $K$ and $K'$ points of the phonon Brillouin zone are degenerate with energy $\hbar \omega_K = 161~{\rm meV}$ and, under the assumption that they are equally populated at the temperature $T_K(t)$, their contributions will be considered together with a singe function ${\cal Q}_K(T_e,T_K)$. These phonon induce inter-valley electronic transitions: the $K$-point phonons scatter electrons from valley $K$ to $K'$, while the $K'$-phonons scatter them in the opposite way. For these phonons,~\cite{Sohier2014}
\be
\big|U_{\lambda,\lambda',K/K'} ({\bm k}, {\bm k}',{\bm q})\big|^2 = g_{K}^2 \big[1 \mp \lambda \lambda' \cos(\varphi_{\bm k} - \varphi_{{\bm k}'})\big]
~.
\ee
In these equations,
\begin{eqnarray}
g_{\alpha} = \beta_\alpha \sqrt{\frac{\hbar}{2 \rho_m \omega_\alpha}}
~,
\end{eqnarray}
where~\cite{Sohier2014} the graphene mass density is $\rho_m = 7.6\times 10^{-7}~{\rm kg/m}^2$. We note that $\beta_K^2= 2 \beta_\Gamma^2$, and that the latter can be rewritten as~\cite{Sohier2014}
\begin{eqnarray}
\beta_\Gamma = \frac{3}{2} \frac{\partial t}{\partial a}
~,
\end{eqnarray}
where $t$ is the nearest-neighbor hopping amplitude and $a$ the carbon-carbon distance. The parameter $\beta_\Gamma$ has been calculated by DFT, GW and also estimated experimentally~\cite{Sohier2014}. Its value varies quite substantially depending on the source. To be conservative, we will use $\beta_\Gamma = 11.4 ~{\rm eV}/$\AA~\cite{Sohier2014} as taken from the GW level of \textit{ab initio} simulations, to be consistent with the chosen Fermi velocity of $v_{\rm F} = 10^6$ m/s. For later use, we define 
\begin{eqnarray}
{\tilde g}_{\alpha} = \beta_\Gamma \sqrt{\frac{\hbar}{2 \rho_m \omega_\alpha}}
~.
\end{eqnarray}

Multiplying Eq.~(\ref{eq:Boltmann_def}) by the energy measured from the chemical potential $\mu \equiv \mu(T_e)$, $\varepsilon_{{\bm k},\lambda} - \mu$, integrating it over momentum ${\bm k}$ and summing over the band $\lambda$, we get the equation of motion for the electron energy, $\partial_t {\cal E}(T_e) = - \sum_\alpha {\cal Q}_\alpha(T_e,T_\alpha)$, where
\begin{eqnarray} \label{eq:Q_rate_gen}
{\cal Q}_\alpha(T_e,T_\alpha) = \hbar \omega_\alpha \frac{4\pi}{\hbar N_{\rm f}} {\tilde g}_\alpha^2 \big[n_\alpha(T_\alpha)  - n_\alpha(T_e) \big]  
\big[{\tilde {\cal Q}}(\mu,\omega_\alpha) - {\tilde {\cal Q}}(\mu+\omega_\alpha,\omega_\alpha) \big]
~,
\end{eqnarray}
is the heat density dissipated in collisions with the phonons of mode $\alpha$. Note that ${\cal Q}_\Gamma(T_e,T_K)$ contains the effect of both longitudinal and transverse phonons, while ${\cal Q}_K(T_e,T_K)$ accounts also for the contribution of modes at $K'$. In Eq.~(\ref{eq:Q_rate_gen}), $N_{\rm f} = 4$ is the spin-valley degeneracy of graphene, and
\begin{eqnarray} \label{eq:tilde_Q_def}
{\tilde {\cal Q}}(\mu,\omega) &=& \int_{-\infty}^{+\infty} d\varepsilon ~\nu(\varepsilon) ~\nu(\varepsilon - \omega) \left[f\left(\frac{\varepsilon - \mu}{k_{\rm B} T_e}\right)  - \Theta(-\varepsilon) \right]
~.
\end{eqnarray} 
In this equation, $\nu(|\varepsilon|) = N_{\rm f}|\varepsilon|/(2\pi \hbar^2 v_{\rm F}^2)$ is the Dirac-fermion density of states. Interestingly, ${\cal Q}_\Gamma$ and ${\cal Q}_K$ have the same functional dependence on the respective phonon frequencies and temperatures, {\it i.e.} all coefficients in the two definitions are the same. This is due to the fact that, while the phonons at $\Gamma$ are doubly degenerate, the interaction vertex $g_K^2$ is twice $g_\Gamma^2$.
From Eq.~(\ref{eq:Q_rate_gen}), using the fact that all phonons of a given mode $\alpha$ have the same frequency, it is possible to define the rate of phonon emission
\begin{eqnarray} \label{eq:rate_phonon_emission_def}
{\cal R}_\alpha(T_e,T_\alpha) = \frac{{\cal Q}_\alpha(T_e,T_\alpha)}{\hbar \omega_\alpha}
~.
\end{eqnarray}

Introducing the (density of) electronic heat capacity 
\begin{eqnarray} \label{eq:C_e_def}
C_e(T_e)= \int_{-\infty}^{\infty} d\varepsilon ~\nu(|\varepsilon|) (\varepsilon - \mu) \left[-\frac{\partial f\big[(\varepsilon-\mu)/(k_{\rm B} T_e)\big]}{\partial \varepsilon}\right] \left[ \frac{\varepsilon-\mu}{T_e} + \frac{\partial \mu}{\partial T_e} \right]
~,
\end{eqnarray}
we finally get
\begin{eqnarray} \label{eq:energy_EOM_2}
C_e(T_e) \partial_t T_{\rm e} = - \sum_\alpha {\cal Q}_\alpha(T_e,T_\alpha)
~,
\end{eqnarray}
from which it is possible to define the instantaneous electron-$\alpha$-phonon cooling time $\tau_\alpha$
\begin{eqnarray} \label{eq:cooling_time}
\tau_\alpha = \left[ \frac{{\cal Q}_\alpha(T_e,T_\alpha)/C_e(T_e)}{T_e - T_\alpha} \right]^{-1}
~.
\end{eqnarray}

In all equations, the chemical potential is determined by (numerically) inverting the equation 
\begin{eqnarray} 
n = \frac{N_{\rm F} (k_{\rm B} T_e)^2}{2\pi (\hbar v_{\rm F})^2} \big[ {\rm Li}_2(-e^{-\mu/(k_{\rm B} T_e)}) - {\rm Li}_2(-e^{\mu/(k_{\rm B} T_e)})\big]
~,
\end{eqnarray} 
fixing the electronics density $n$. Here, ${\rm Li}_n(x)$ is the $n$-th polylogarithmic function. The derivative of the chemical potential is readily evaluated, under the assumption of a constant density, as
\begin{eqnarray}
\frac{\partial \mu}{\partial T_e} =
\frac{\mu}{T_e} - k_{\rm B} \frac{{\rm Li}_2(-e^{-\mu/(k_{\rm B} T_e)}) - {\rm Li}_2(-e^{\mu/(k_{\rm B} T_e)})}{\ln(1+e^{\mu/(k_{\rm B}T_e)})+\ln(1+e^{-\mu/(k_{\rm B}T_e)})}
~.
\end{eqnarray}

We note that the integral in Eq.~(\ref{eq:C_e_def}) can be carried out analytically to give
\begin{eqnarray} \label{eq:C_final}
C_e(T_e) &=& k_{\rm B} \frac{N_{\rm F} (k_{\rm B}T_e)^2}{2\pi (\hbar v_{\rm F})^2} \Big\{ 
\left[
3 \big(F_{+,2}(x) - F_{-,2}(x)\big) - 4 x \big(F_{+,1}(x) - F_{-,1}(x)\big) + x^2 \big(F_{+,0}(x) - F_{-,0}(x)\big)
\right]
\nonumber\\
&+&
\frac{\partial \mu}{\partial (k_{\rm B} T)}
\left[
2 \big(F_{+,1}(x) - F_{-,1}(x)\big) - x \big(F_{+,0}(x) - F_{-,0}(x)\big)
\right]
\Big\}_{x = \mu/(k_{\rm B} T_e)}
~.
\end{eqnarray}
Here we defined
\begin{eqnarray}
&&
F_{-,n}(x) = \int_{-\infty}^{0} dy~y^n \big[f\left(y - x\right) - \Theta(-y)\big] 
~,
\nonumber\\
&&
F_{+,n}(x) = \int_{0}^{+\infty} dy~y^n \big[f\left(y - x\right) - \Theta(-y)\big] 
~.
\end{eqnarray}
These integrals are calculated explicitly for the first few values of $n$ to give
\begin{eqnarray}
&&
F_{\pm,0}(x) = \pm \ln(1+e^{\pm x})
~,
\nonumber\\
&&
F_{\pm,1}(x) = - {\rm Li}_2(- e^{\pm x})
~,
\nonumber\\
&&
F_{\pm,2}(x) = \mp 2 {\rm Li}_3(- e^{\pm x})
~,
\nonumber\\
&&
F_{\pm,3}(x) = - 6 {\rm Li}_4(- e^{\pm x})
~.
\end{eqnarray}

Similarly, defining
\begin{eqnarray}
&&
{\tilde F}_{n}(x,w) = \int_{0}^{w} dy~y^n \big[f\left(y - x\right) - \Theta(-y)\big] 
~.
\end{eqnarray}
such that
\begin{eqnarray}
&&
{\tilde F}_{0}(x,w) = \ln(1+e^x) - \ln(1+e^{x-w})
~,
\nonumber\\
&&
{\tilde F}_{1}(x,w) = {\rm Li}_2(- e^{x-w}) -  {\rm Li}_2(- e^{x}) - w \ln(1+e^{x-w})
~,
\nonumber\\
&&
{\tilde F}_{2}(x,w) = 2 {\rm Li}_3(- e^{x-w}) - 2 {\rm Li}_3(- e^{x}) + 2 w {\rm Li}_2(- e^{x-w}) - w^2 \ln(1+e^{x-w})
~,
\end{eqnarray}
Eq.~(\ref{eq:tilde_Q_def}) can be integrated analytically to give
\begin{eqnarray}
{\tilde {\cal Q}}(\mu,\omega) 
&=&
\left(\frac{N_{\rm F}}{2\pi(\hbar v_{\rm F})^2}\right)^2 (k_{\rm B}T_e)^3 
\Big\{
\big[F_{+,2}(x) + F_{-,2}(x)\big] - w \big[F_{+,1}(x) + F_{+,1}(x)\big]
\nonumber\\
&-&
2 \big[ {\tilde F}_{2}(x,w) - w {\tilde F}_{1}(x,w)\big]
\Big\}_{x=\mu/(k_{\rm B} T_e), w = \omega/(k_{\rm B T_e})}
~.
\end{eqnarray} 

\subsection{The phonon density of states}
Given the temperatures and densities we are working at, the largest contribution to electron cooling comes from intraband processes. These must satisfy:
\be
v_{\rm F} |{\bm k}+{\bm q}| = v_{\rm F} k -\omega
~\Rightarrow~
\left\{
\begin{array}{l}
	0 < \omega < v_{\rm F} k
	\\
	{\displaystyle q_\pm(\theta) = - k \cos(\theta) + \sqrt{k^2 \cos^2(\theta)+\frac{\omega^2}{v_{\rm F}^2} - 2\frac{\omega}{v_{\rm F}} k} }
\end{array}
\right.
~.
\ee
The maximum and minimum momenta of the phonon are
\ber \label{eq:q_min_max_defs}
&& q_{\rm min} = q_-(\pi) = \omega/v_{\rm F}
~,
\nn\\
&& q_{\rm max} = q_+(\pi) = 2 k - \omega/v_{\rm F}
~.
\eer

The maximum and minimum values of $k$ are determined by considering the difference of Fermi functions:
\begin{equation}
f\left(\frac{\varepsilon_{{\bm k},+}-\mu}{k_{\rm B}T}\right) - f\left(\frac{\varepsilon_{{\bm k},+}-\mu-\hbar\omega}{k_{\rm B}T}\right)
~.
\end{equation}
The function ${\cal F}(x,y) = f(x-y)-f(x)$, where $x=(\varepsilon_{{\bm k},+}-\mu)/(k_{\rm B}T)$ and $y=\hbar\omega/(k_{\rm B} T)$ can be viewed as a distribution in $x$. Its mean value is approximately $y/2$, while its standard deviation is 
\begin{equation}
\sigma(y) = \sqrt{\frac{4\pi^2+y^2}{12}}
~.
\end{equation}]
Therefore, the difference in Fermi functions implies that $k$ is approximately bounded within
\begin{equation}\label{eq:k_boundaries}
\mu+\frac{\hbar \omega}{2}- \nu k_{\rm B} T \sigma\left(\frac{\hbar \omega}{k_{\rm B}T}\right) \lesssim v_{\rm F} k \lesssim \mu+\frac{\hbar \omega}{2}+\nu k_{\rm B} T \sigma\left(\frac{\hbar \omega}{k_{\rm B}T}\right)
~,
\end{equation}
where $\nu$ here is a fitting parameter of order one. Therefore,
\begin{equation}
\frac{\omega}{v_{\rm F}} \lesssim q \lesssim 2 \left[ \frac{\mu}{\hbar v_{\rm F}} + \nu \frac{k_{\rm B} T}{\hbar v_{\rm F}} \sigma\left(\frac{\hbar \omega}{k_{\rm B}T}\right)\right]
~.
\end{equation}
To obtain this equation we substituted the maximum $k$ of Eq.~(\ref{eq:k_boundaries}) into $q_{\rm max}$ of Eq.~(\ref{eq:q_min_max_defs}). We then get
\begin{equation}
\left\{
\begin{array}{l}
	{\displaystyle \varepsilon_{\rm min}(T_e) = \hbar \omega}
	\vspace{0.2cm}\\
	{\displaystyle \varepsilon_{\rm max}(T_e) = 2 \left[ \mu + \nu k_{\rm B} T \sigma\left(\frac{\hbar \omega}{k_{\rm B}T}\right)\right]}
\end{array}
\right.
~.
\end{equation}

\newpage
%\bibliography{WSe2GrapheneCooling_biblio4}
\twocolumngrid
\appendix

\end{document}